\documentclass[reprint,aip,rsi,amsmath,amssymb,nofootinbib]{revtex4-1}

\usepackage{graphicx}% Include figure files
\usepackage{dcolumn}% Align table columns on decimal point
\usepackage{bm}% bold math
\usepackage{float}
\usepackage{caption}
\usepackage{subcaption}
\usepackage{xcolor}
\usepackage{soul}
\usepackage{appendix}
\usepackage{chngcntr}
\usepackage{multirow}
\usepackage{textcomp, gensymb}
\usepackage{comment}
\usepackage{soul}

\begin{document}
\setstcolor{red}
%\title{Understanding Moderate Pressure Capacitively Coupled Radiofrequency Plasma-assisted Synthesis of Carbon Nanomaterials: Reactive Species and Gas Heating}

\title{Importance of Gas Heating in Capacitively Coupled Radiofrequency Plasma-assisted Synthesis of Carbon Nanomaterials}

\author{Tanvi Nikhar$^*$}
%\email{nikharta@msu.edu}
\affiliation{Department of Electrical and Computer Engineering, Michigan State University, MI 48824, USA}
%\thanks{These authors contributed equally to this work} 

\author{Sankhadeep Basu$^*$}
%\email{basusank@msu.edu}
\affiliation{Department of Chemical Engineering and Material Science, Michigan State University, MI 48824, USA}
\affiliation{Department of Mechanical Engineering, Michigan State University, MI 48824, USA}
%\thanks{These authors contributed equally to this work}

\def\thefootnote{*}\footnotetext{These authors contributed equally to this work}

\author{Shota Abe}
%\email{sabe@pppl.gov}
\affiliation{Discovery Plasma Science Department, Princeton Plasma Physics Laboratory, NJ 08542, USA}

\author{Shurik Yatom}
%\email{syatom@pppl.gov}
\affiliation{Discovery Plasma Science Department, Princeton Plasma Physics Laboratory, NJ 08542, USA}

\author{Yevgeny Raitses}
%\email{yraitses@pppl.gov}
\affiliation{Discovery Plasma Science Department, Princeton Plasma Physics Laboratory, NJ 08542, USA}

\author{Rebecca Anthony}
\email{ranthony@msu.edu}
\affiliation{Department of Chemical Engineering and Material Science, Michigan State University, MI 48824, USA}
\affiliation{Department of Mechanical Engineering, Michigan State University, MI 48824, USA}

\author{Sergey V. Baryshev}
\email{serbar@msu.edu}
\affiliation{Department of Electrical and Computer Engineering, Michigan State University, MI 48824, USA}
\affiliation{Department of Chemical Engineering and Material Science, Michigan State University, MI 48824, USA}

\begin{abstract}
%Carbon’s dual hybridization routes ($sp^2$  and $sp^3$) leads to hugely disparate properties for the resulting allotropes, ranging from the unique band structure and electrical conductivity of graphene to the preeminent combined electrical, thermal, and structural properties of diamond.

In pursuit of diamond nanoparticles, a capacitively-coupled radio frequency (CCRF) flow-through plasma reactor was operated with methane-argon gas mixtures. Signatures of the final product obtained microscopically and spectroscopically indicated that the product was an amorphous form of graphite. %, i.e. an $sp^2$ hybridized allotrope of carbon.
This result was consistent irrespective of combinations of the macroscopic reactor settings. %To gain a fundamental understanding of gas kinetics and 
To explain the observed synthesis output, measurements of C$_2$ and gas properties were carried out %molecular species and densities, their distributions, and the gas temperature were conducted
by %planar 
laser-induced fluorescence and optical emission spectroscopy.
%These diagnostics revealed first that the abundance of C$_2$ was as high as $\sim 10^{12}$ cm$^{-3}$.
%Second and more
 Strikingly, the results indicated a strong gas temperature gradient of %150 K 
 100 K per mm from the center of the reactor to the wall. Based on additional plasma imaging, a model of hot constricted region (filamentation region) was then formulated. It illustrated that, while the hot constricted region was present, the bulk of the gas was not hot enough to facilitate diamond $sp^3$ formation: %because rf power was consumed via heating a narrow filament-like region in the center of the reactor, also explaining the thermal gradient as found from PLIF and HROES.
 characterized by much lower reaction rates, when compared to $sp^2$, $sp^3$ formation kinetics are expected to become exponentially slow.
 %in the gas only slightly warmer than the room temperature. While the hot constricted region is present, the overall gas temperature favors formation of $sp^2$.
 This result was further confirmed by experiments under identical conditions but with a H$_2$/CH$_4$ mixture, where no output material was detected: if graphitic $sp^2$ formation was expected as the main output material from the methane feedstock, atomic hydrogen would then be expected to etch it away in situ, such that the net production of that $sp^2$-hybridized solid material is nearly a zero.

%$E_\text{a}$
%H$_\text{2}$/CH$_\text{4}$ 
%$\sim1000-1300$ K

\end{abstract}

\maketitle

\section{\label{sec:Introduction}Introduction}
Low-temperature plasmas (LTP) have been an enabling technology and indispensable tool for the microelectronics and semiconductor industry.\cite{graves2023} The critical importance of LTP technologies keeps growing in the modern era of sub-10 nm node 3D integration, where UV lithography, etching, passivation and ultrathin film growth are the most critical steps of optoelectronic integrated circuits fabrication and integration. At the same time, LTP-based syntheses occupy a special place in semiconductor nanomaterial manufacturing which is important in photovoltaics, optoelectronics, and quantum information sciences.\cite{hori2017, kortshagen2016} Among other types of LTP reactors, sub-atmospheric capacitively coupled RF continuous flow-through reactors gained a lot of attention and implementations since the early 2000s due to their simplicity, high production rates/material yields, simple means of material collection and post-synthesis treatment and integration (including 3D printing).\cite{anthony2009, kortshagen2016_springer, mangolini2007, woodard2018, exarhos2018, thimsen2015, uner2019, izadi2019, mandal2018, ho2021, dsouza2023, askari2015, nozaki2007} By now, flow- through CCRF LTP reactors succeeded in creating a profound list of compound (oxide, sulfide, nitride, and phosphide) and elemental group IV semiconductors.\cite{kortshagen2016} To date, syntheses of Group IV nanomaterials have included well-established and vetted synthetic routes for Si, Ge, and SiGe nanocrystals\cite{mangolini2005, pi2009, gresback2007}, SiC, and graphite or $sp^2$ moieties of carbon, but never diamond itself. 

Historically, microwave plasma-assisted CVD is used to deposit diamonds in bulk thin film form where substrate surface kinetics governs the growth. Gas phase diamond nucleation and growth is a subject that was studied much less. In the 1970s, Deryagin $et$ $al.$ proposed the possibility of nucleating diamond in the gas phase inside a plasma.\cite{deryagin1979} Experimental work on gas-phase nucleation was conducted by Frenklach $et$ $al.$ at Penn State using a low-pressure microwave plasma. As-synthesized materials had to be treated and purified before sub-micron diamond crystals could be retrieved thereby proving %Fedoseev’s
Deryagin's hypothesis.\cite{frenklach1989} At that time, there were a limited number of reports corroborating the results by Frenklach and co-workers. Much later, Sankaran $et$ $al.$ produced nano-diamonds using a dc atmospheric pressure flow-through micro-discharge plasma.\cite{kumar2013} Synthesis of diamond nanocrystals in CCRF flow-through reactors remains elusive. Given the scarcity of characterization results of the plasmas that successfully produced nano-diamonds, the conditions required for nanocrystals of $sp^3$-bonded carbon (i.e. nano-diamond) synthesis are not well known.

At face value, this synthesis route to diamond nanocrystals appears obvious, when compared to that of silicon nanocrystals. If diamond cubic Si nanocrystals are synthesized in SiH$_4$/Ar plasmas, then simple switching to CH$_4$/Ar plasma would be expected to produce diamond cubic diamond nanocrystals. Such a switching was successfully implemented in CVD reactors at the Argonne National Lab where ultra-nano-crystalline-diamond (UNCD) was patented and trademarked.\cite{auciello2022} In this work, we study this trivial switching in a moderate pressure flow through CCRF reactor and find that carbon allotropy plays the critical role in engineering plasma and gas phase conditions to successful diamond nano-crystal synthesis by LTP flow through reactors.

 %Frenklach $et$ $al.$ \cite{Frenklach1989}  

\section{\label{sec:Experiment}Experiment}

%\subsection{Reactor Setup and Sample Collection}
\subsection{\label{Setup}Reactor Setup and Sample Characterization}
\begin{figure}[h]
\centering
    \includegraphics[width=0.35\textwidth]{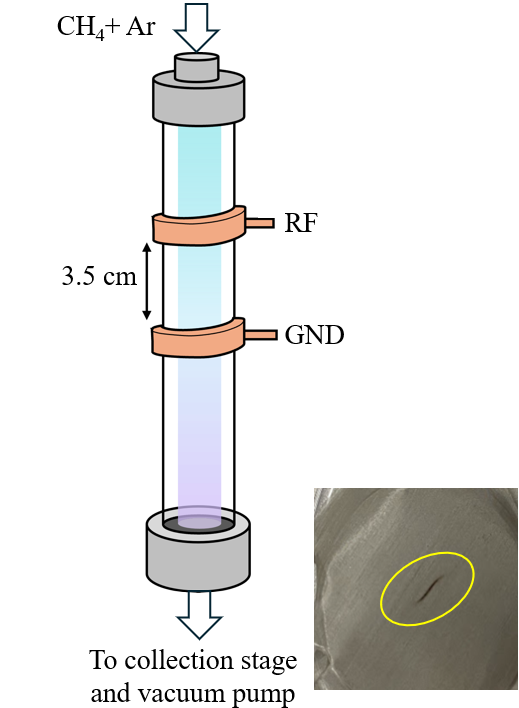}
    \caption{Schematic of the reactor with the picture of carbon nanoparticle deposition on stainless steel mesh (orifice-shaped deposition area marked by the yellow oval).}
    %Schematic of reactor. Photos of carbon nanoparticle deposition on stainless steel meshes at 10sccm and 2 sccm are shown. Yellow ovals mark the orifice-shaped deposition areas.
\label{fig:schematic_reactor}
\end{figure}

The syntheses were carried out in a flow-through tubular reactor. A schematic of the reactor is shown in Fig. \ref{fig:schematic_reactor}. The gases used for the synthesis were argon (or hydrogen) and methane, where methane was the carbon precursor for the reaction. The reactor consisted of a 1.27 cm outer diameter quartz cylindrical tube of length 23.5 cm to which the gas mixture of 100 standard cubic centimeters per minute (sccm) of argon (or hydrogen) and 2 sccm of methane flowed via mass flow meters (Alicat Inc.). %The argon (or hydrogen) flow rate was kept constant at 100 standard cubic centimeters per minute (sccm) and the methane flow rate was changed in different experimental runs. For this work, methane flow rates of 10 sccm and 2 sccm are directly compared.  
The total pressure measured upstream of the reaction region was between 4-4.2 Torr. RF power of 200 W at 13.56 MHz from the power supply (AG0313, T{\&}C Power Conversion) was delivered to the reactor via a copper ring electrode placed around the tube, with a ground electrode placed 3.5 cm downstream of the powered electrode. An impedance matching network (MFJ 989D) was used in addition to the power supply to minimize the reflected power. %Nanomaterials were collected on stainless steel meshes or Si coupons placed downstream of the reactor.

Nanomaterials were collected on stainless steel meshes or Si coupons placed downstream of the reactor. Samples were characterized using scanning electron microscopy (SEM) on a Verios 460XHR instrument%;  transmission electron microscopy (TEM) on a JEOL 2200FS microscope,
and visible (532 nm) and UV (325 nm) Raman spectroscopy with Horiba LabRAM ARAMIS Raman instrument.

%In this work, the laser-induced fluorescence (LIF) approach is used to measure the density of C$_2$ molecules, as one of the suspected pre-cursors for the synthesis of carbon nanomaterials.\cite{yatom2018_MRS, yatom2017_carbon, stratton2018} The LIF approach is a versatile tool for the absolute density measurement of atoms and molecules, including C$_2$.\cite{vekselman2018} 

%\subsection{Sample characterization}
%Samples were characterized using scanning electron microscopy (SEM) on a Verios 460XHR instrument%;  transmission electron microscopy (TEM) on a JEOL 2200FS microscope, and visible (532 nm) and UV (325 nm) Raman spectroscopy with Horiba LabRAM ARAMIS Raman instrument.

\subsection{Laser Induced Fluorescence (LIF)}
LIF was used to measure the absolute densities of carbon dimer C$_2$ as one of the suspected pre-cursors for the synthesis of carbon nanomaterials generated from the plasma-induced reactions inside the reactor.\cite{yatom2018_MRS, yatom2017_carbon, stratton2018} The LIF approach is a versatile tool for the absolute density measurement of atoms and molecules, including C$_2$.\cite{vekselman2018} An Nd: YAG+OPO tunable laser system\cite{yatom2022_physD, yatom2023_plasma} was used to generate the laser pulse at 438.8 nm, to excite the (2,0) swan band transitions, and the fluorescence from the corresponding excitation was detected around 470 nm. The excitation beam was generated by a Nd: YAG laser producing nanosecond duration laser pulses with a frequency of 20 Hz. The laser beam is shaped with a cylindrical lens into a laser sheath with a height of 4.2 mm and width of 754 $\mu$m. 

The laser energy was limited to 650 $\mu$J per pulse. The beam passed through the cylindrical tube from one end to the other and was terminated with the help of a beam dump placed at the other side of the reactor (see LIF schematic). The fluorescence signal is observed in a direction normal to the incoming excitation beam. A mechanical slit was situated on the detection axis, with its position and width adjusted to eliminate reflections from the sides of the tube. The detector (iCCD camera PiMAX4, Princeton Instruments, Gen III HBf intensifier) was used with a spectral filter (470 nm, FWHM 10 nm). The camera gate and the laser pulse are synchronized with the help of a pulse delay generator (BNC 575-4C). %To track the C$_2$ radical densities along the length of the tube, 3 different locations were selected. These were a) 3 cm above the powered electrode (location 1), b) 2 cm below the ground electrode (location 2), and c) 6 cm below the ground electrode (location 3). %At every location, 2 different flow rates of methane (2 sccm and 10 sccm) were studied while keeping other variables constant.

The determination of the absolute C$_2$ molecular density from the LIF signal followed the protocol developed by Luque $et$ $al.$\cite{luque1997} The instrumental calibration has been done by Rayleigh scattering, with the procedure described in detail in Refs. \onlinecite{yatom2022_physD, yatom2023_plasma}.

\subsection{Optical emission spectroscopy (OES)}
Broadband OES was performed using Ocean Optics HR2000+ spectrometer for preliminary detection of various species present in the plasma. For the measurement of gas temperature, high-resolution OES (HROES) was performed using (HRS SpectraPro 750 monochromator coupled with PiMAX4 iCCD camera) to detect CH(A-X) and C$_2$ emissions. In the HROES setup the whole length of the discharge tube was imaged (with demagnification) on the spectrometer slith (14 mm), so that the obtained spectra contain spatially resolved emission from  all the investigated locations along the vertical axis of the iCCD screen, while the horizontal axis corresponds to the dispersion axis.  A similar approach was utilized in earlier work on plasma characterization.\cite{yatom2020_phychem}

\section{\label{sec:Results}Results}
\begin{figure}
\centering
    \includegraphics[width=0.5\textwidth]{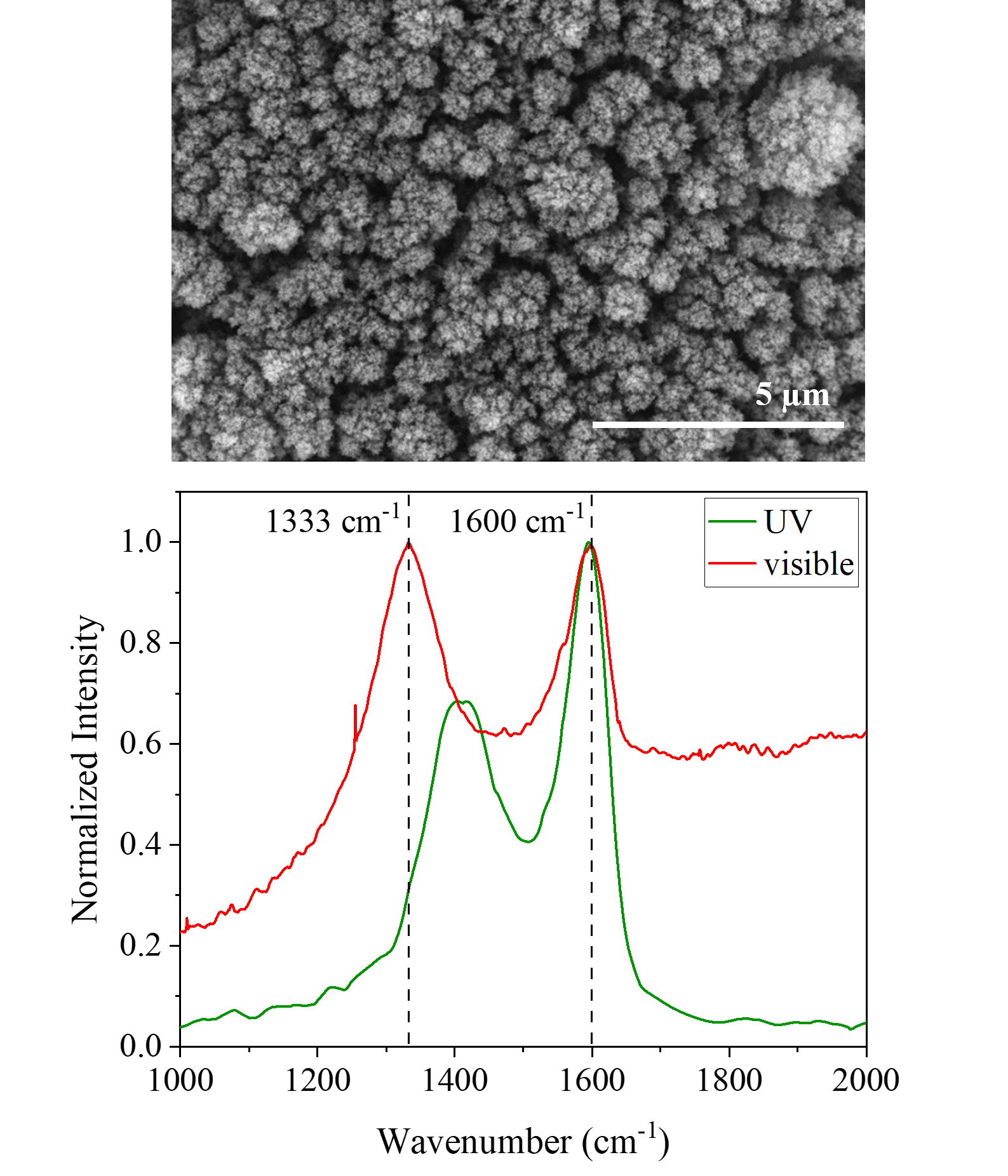}
     \caption{(Top) SEM image and (bottom) Raman spectra of the synthesized material.}
    %\caption{SEM and Raman of the synthesized materials. The SEM image is a representative from the 2 sccm methane flowrate condition; however the findings from SEM are similar for both.}
\label{fig:SEM_&_Raman}
\end{figure}

%For the present study, the Ar flowrate and reactor pressure were kept constant while the methane flowrate was moderated to be either “low” (2 sccm) or “high” (10 sccm). 
%The input RF power generates the plasma and breaks down the incoming gas mixture into different energetic radicals, along with the production of electrons and ions. These radicals undergo numerous chemical reactions that ultimately result in nanomaterial synthesis.
Fig. \ref{fig:SEM_&_Raman} (top) shows the SEM image of the synthesized nanomaterial exhibiting agglomerated carbon colonies indicating its amorphous nature. The Raman spectra from Fig. \ref{fig:SEM_&_Raman} (bottom) shows broad D and G bands from the visible Raman typical to $sp^2$ bonded carbon. The absence of a 1333 cm$^{-1}$ peak in the UV Raman spectrum indicates the lack of any $sp^3$ bonded carbon in the sample. These characterization findings raise the question of why crystalline carbon allotropes like graphite, diamond, or lonsdaleite were not created within the reactor parameters studied here. With Si or other materials, the primary variable responsible for crystallization is the temperature of the growing nanomaterial environment (as realized via the delivered power to the reactor),\cite{anthony2009, galli2009, mangolini2009, lopez2014} and so it is important to get an estimate of the gas temperature inside the reactor along with the reactive species responsible for the formation of the obtained amorphous $sp^2$ carbon.

\begin{figure}[h]
\centering
    \includegraphics[width=0.5\textwidth]{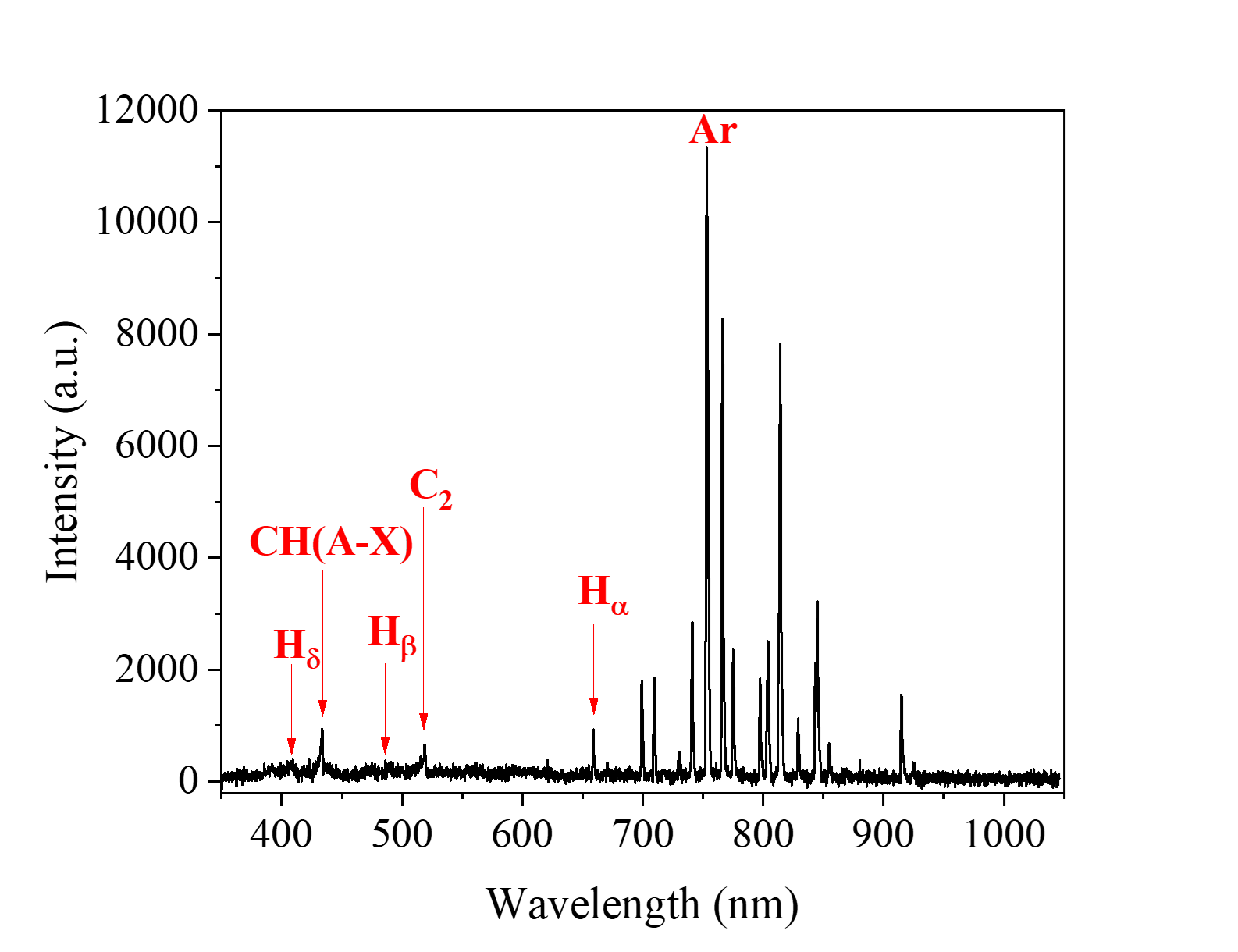}
    \caption{Low-resolution OES spectrum showing the main plasma-activated species in the Ar/CH$_4$ plasma.} %(Low-resolution OES spectrum showing the main plasma-activated species (Ar flowrate 100 sccm, CH4 flowrate XXX}
\label{fig:low-res_OES}
\end{figure}

To identify the reactive species present in our reactor, we carried out a preliminary broadband OES, which revealed the presence of C$_2$, CH, H, and Ar lines. The OES spectrum for the Ar/CH$_4$ plasma is shown in Fig. \ref{fig:low-res_OES}.
%A representative OES spectrum for the condition of (Low/high) methane flow rate is shown in Fig.2.

%Because there is no %other 
%carbon source other than methane, it can be safely said that C$_2$ and CH are a result of methane breakdown. In plasma systems involving carbon, C$_2$ has been said to play a major role in the final synthesis product and hence it became important to get a quantitative idea about its density.\cite{mahoney2019} From the swan band transitions in C$_2$,  

Since C$_2$ is attributed to play a major role in $sp^2$ carbon formation,\cite{mahoney2019} LIF was performed using the swan band transitions in C$_2$ to quantify its number density in our reactor. The LIF signal $S_F$ is related to the total number density $n_0$ (m$^{-3}$) of the target species by the following equation:\cite{luque1997}

\begin{equation}
S_F  = n_0 f_b \frac{B}{c} E_L \frac{\Gamma}{\Delta \nu} \frac{\tau_eff}{\tau_0} F_{fl} \bigg(\frac{\Omega}{4 \pi} \varepsilon \eta \frac{V}{A_L}\bigg)
\label{eq:S_f}
\end{equation} 
%where $f_b$ = 0.00805 
where $f_b$ = 0.00655 is the Boltzmann factor calculated following the method stated by Luque et al.\cite{luque1997_optica} using the rotational and vibrational temperature of 
%1200 K and 2000 K, respectively (obtained from LIFBASE using CH emission spectrum)} from the following relation:
825 K and 2827 K, respectively corresponding to the C$_2$ emission spectrum(discussed later in this section) from the following relation:

\begin{equation}
    f_b = \frac{exp\bigg(\frac{-E_{vib}}{kT_{vib}}\bigg)(2J+1)exp\bigg(\frac{-E_{rot}}{kT_{rot}}\bigg)}{Q_{vib}Q_{rot}Q_{elec}}
\end{equation}
where $E_{vib}$ and $E_{rot}$ are the vibrational and rotational energies, $(2J+1)$ is the rotational degeneracy. $Q_{vib}$, $Q_{rot}$, and $Q_{elec}$ are the vibrational, rotational and electronic partition functions, respectively. The energies of the rotational and vibration states are calculated from $E_{rot}(J)=bJ(J+1)$,\cite{hollas2004} where $b$ is the effective rotational constant and $J$ is the rotational energy level; $E_{vib}=w_e\bigg(v+\frac{1}{2}\bigg)-w_ex_e\bigg(v+\frac{1}{2}\bigg)^2$ where $v$ is the quantum energy level, $w_e=1641.35$ cm$^{-1}$ is the vibrational wavenumber and $w_ex_e=11.67$ cm$^{-1}$ is the anharmonic constant. Taking into account the vibrational dependence of the rotational constant, $b$ is further expressed as $b=b_e-\alpha\bigg(v+\frac{1}{2}\bigg)$ where $\alpha=0.1661$ cm$^{-1}$ is the vibration-rotation interaction constant and $b_e=1.6324$ cm$^{-1}$ is the rotational constant of the equilibrium state. All the constants used in the calculations have been taken from Ref. \onlinecite{huber1979}.

$B=2.5\times10^9$ m$^2$J$^{-1}$s$^{-1}$  is the absorption coefficient for laser excited rotational transition as given in Ref. \onlinecite{luque1997}; $c=3\times10^8$ ms$^{-1}$ is the speed of light; $E_L=650\times10^{-6}$ J is the laser energy; $\Gamma=1.75$ is the line shape overlap integral obtained using the process stated in Ref. \onlinecite{partridge1995}; $\Delta\nu=504$ m$^{-1}$ is the laser bandwidth; $\tau_{eff}/\tau_0$ is the fluorescence quantum yield, where $\tau_{eff}$ is the effective lifetime shortened by quenching calculated from the time-resolved exponential decay of the LIF signal using Ref. \onlinecite{luque1997} and $\tau_0=100$ ns is the radiative lifetime value corresponding to C$_2$ as given in Ref. \onlinecite{luque1997}. $F_{fl}=1$ is the fraction of light emitted by the pumped transition in the collected spectral region. $\Omega$ is the solid angle of the laser-probed volume $V$. $\varepsilon$ is the transmission efficiency of the optics, $\eta$ is the photoelectric efficiency of the photomultiplier, and $A_L$ is the laser cross-section area. 

The term $\bigg(\frac{\Omega}{4\pi} \varepsilon \eta \frac{V}{A_L}\bigg)$ is determined by the calibration measurements of Rayleigh scattering of air. The Rayleigh signal of the incident light collected at a normal angle is given by the following relation:\cite{luque1997}

\begin{equation}
S_R  = N \frac{E_L}{h c \nu} \frac{\partial \sigma}{\partial \Omega} \bigg(\Omega \varepsilon \eta \frac{V}{A_L}\bigg)
\label{eq:S_R}
\end{equation} 
where $N=p/kT$ is the gas number density, pressure  $p=1$ atm or 101325 Pa, $k=1.38\times10^{-23}$ JK$^{-1}$ is the Boltzmann constant, and gas temperature $T=293$ K; $h=6.626\times10^{-34}$ Js is the Planck's constant; $\nu=(438\times10^{-9})^{-1}$ m$^{-1}$ is the wavenumber of the scattered photon; $\frac{\partial \sigma}{\partial \Omega}=1.418\times10^{-31}$ m$^2$sr$^{-1}$ is the Rayleigh cross section corresponding to the polarizability of air$=1.8\times10^{-30}$ m$^3$ and depolarization ratio of air$=0.028$ calculated using Ref. \onlinecite{reichardt2003}. %[\hl{10.1364/AO.42.004909}] 
The slope of the plot $S_R$ vs ($E_L\times p$), obtained by varying $E_L$ from 1.7 to 17 mJ at atmospheric pressure, is used to calculate the term $\bigg(\frac{\Omega}{4\pi} \varepsilon \eta \frac{V}{A_L}\bigg)$ which is plugged in Eq. \ref{eq:S_f} to give the absolute number density of C$_2$ $\sim10^{18}$ m$^{-3}$. %We hypothesize that this density number is sufficient for the formation of long-range carbon chains %which 
%that nucleate out from the gas phase %upon supersaturation 
%to form $sp^2$ phase carbon nanomaterials.
This concentration of $C_2$ in the highly collisional plasma is sufficient for the formation of $sp^2$ phase carbon nanomaterial.\cite{luque1997, kruis1994}

%It was also found that the variation of methane flowrate resulted in different mass production; the lower flowrate yielded much lesser sample (Figure). Under the SEM microscope, however, both low-methane and high-methane samples exhibited agglomerated carbon colonies. Further examination by TEM revealed that the samples lacked long range order and were mostly amorphous. No lattice fringes were seen, nor any diffraction rings in selected area electron diffraction (SAED). A second confirmation for the amorphous nature came from the UV and visible Raman spectroscopy of the samples. The visible Raman spectra for samples made under both conditions exhibited a broad D peak around 1332 cm$^{-1}$ and a similar G peak around 1600 cm$^{-1}$, characteristic of amorphous carbons. The visible Raman is sensitive to $sp^2$ bonded carbon;  hence to observe the contribution of $sp^3$ bonded carbon, UV Raman was also carried out. As before, it showed D and G peaks with D peak shifted to 1430 cm$^{-1}$ (?) for samples from both flow conditions. The intensity of the G peak however was higher signifying the $sp^2$ nature of the samples.

\begin{figure}
\centering
    \includegraphics[width=0.5\textwidth]{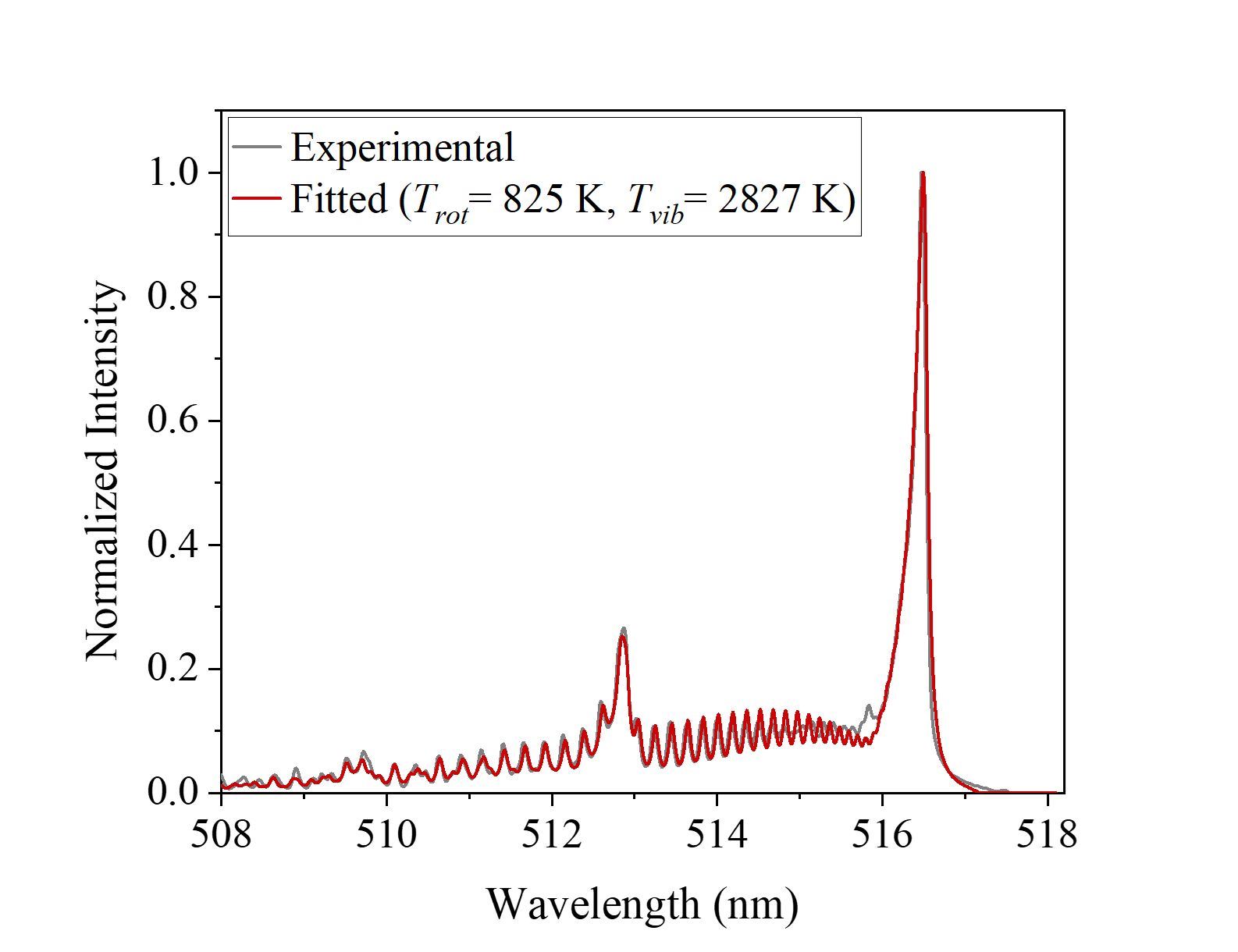}
    \caption{Experimental (from HROES) vs fitted spectrum for C$_2$ used for gas temperature calculation.}
    
    %The ccrf setup at PCRF for LIF and HROES measurements, under conditions of XXX sccm methane flowrate. (right) Representative HROES spectrum for C2 used for gas temperature modelling.
\label{fig:HROES}
\end{figure}

%High-resolution optical emission spectroscopy (HROES) of C$_2$ and CH(A-X) was explored as a means to estimate the temperature in the reactor during synthesis. 
To estimate the gas temperature in the reactor during synthesis, HROES was used to measure the C$_2$ emission spectrum around 516 nm and was fitted in Specair  
software\cite{specair} to give a rotational temperature of 825 K and a vibrational temperature of 2827 K (see Fig. \ref{fig:HROES}). To extrapolate the reactor gas temperature from this measurement requires the assumption that there is thermal equilibrium among the heavy species in the plasma (though the electrons have much higher energies), and that the rotational temperature is similar to the gas temperature.\cite{bruggeman2014} %From these assumptions, the C$_2$  spectrum indicated the gas temperature in the reactor is $\sim800-900$ K. 
From these assumptions, the C$_2$  spectrum indicated the gas temperature of $\sim800-900$ K in the reactor.
%The CH(A-X) emission spectrum around 430 nm was fitted in LIFBASE Spectroscopy Tool\cite{lifbase} %[\hl{LIFBASE spectroscopy tool (2021) https://www.sri.com/case-studies/lifbase-spectroscopy-tool/}] 
%to give a higher rotational temperature of 1200 K and a vibrational temperature of 1200 K. However, this temperature is still 
This temperature is too low to induce crystallization via gas heating because crystallization temperatures of graphite or diamond exceed 1300 K. Furthermore, in a previous study, the authors reported that higher temperature promotes graphitization in CVD systems.\cite{nikhar2020} 

The gas temperature finding of $\sim800-900$ K deserves closer consideration. In capacitively coupled flow-through reactors, gas temperature has been said to remain close to room temperature, exceeded only by a couple of hundred Kelvin.\cite{agarwal2012} In the present study, the variable in question is apparently high in the traditional sense. Outside wall temperature measured by an attached thermocouple indicates that there must be a temperature gradient from the reaction zone (high temperature) to the wall (low temperature).

\section{\label{sec:Discussion}Discussion}
The most plausible hypothesis that could explain self-consistently the extreme thermal gradient in the reactor (%$\sim600$ 
$\sim500$ K between centerline and wall) and, otherwise the overall amorphous $sp^2$ product output is the presence of a hot filamentary region (in other words, plasma constriction) in the centerline of the reactor. If this were true, then RF power would be absorbed by the constricted plasma, thereby immensely increasing the power density and hence heating it to the condition of a hot wire. The rest of the reactor where gas flows at a given velocity would then receive additional heating through thermal conductivity and diffusivity.

\begin{figure}[t]
\centering
    \includegraphics[width=0.35\textwidth]{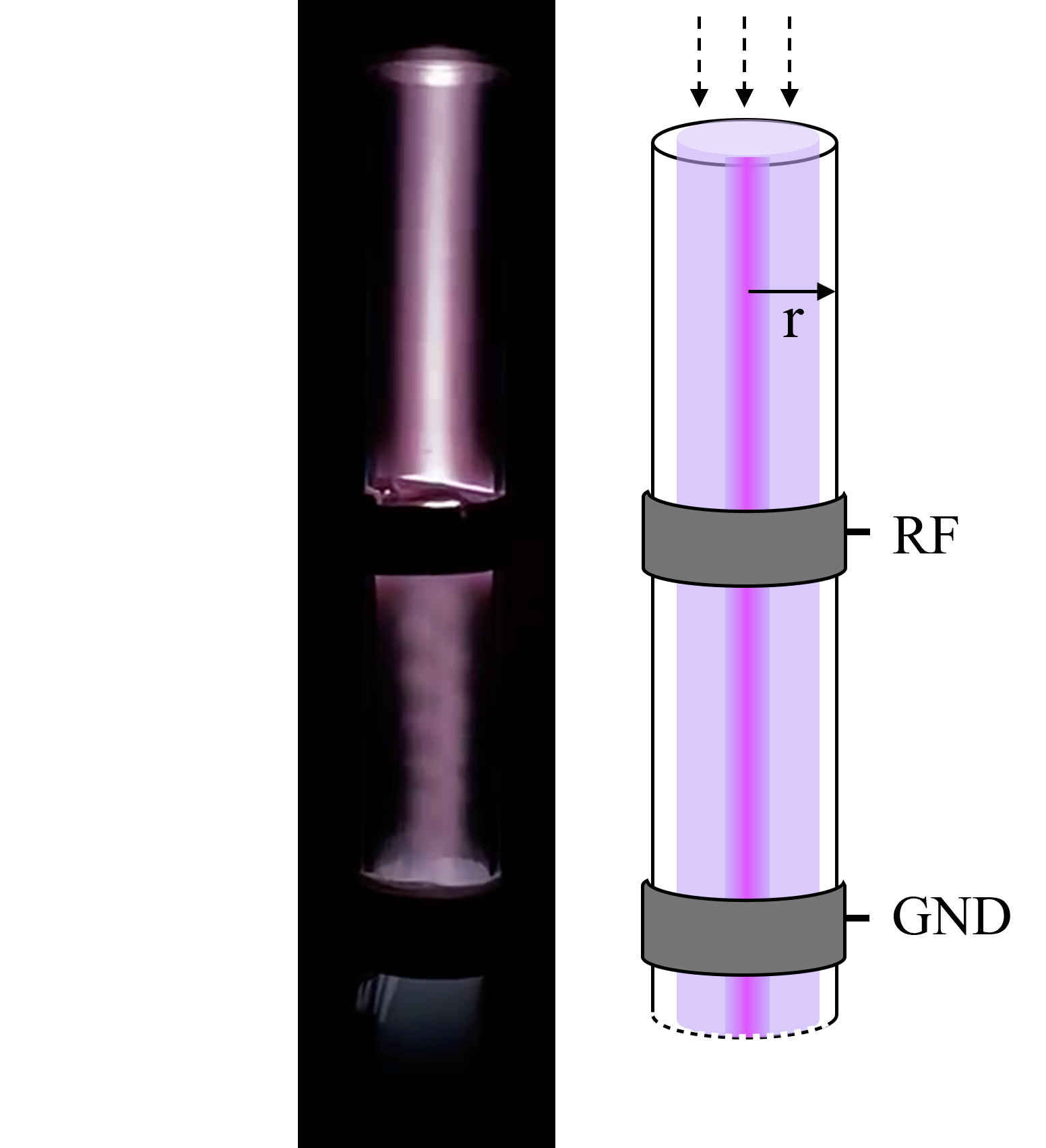}
    \caption{(Left) Photograph and (right) demonstration and model of the hot filament.}
    %Photograph of hot filament (left) demonstration and model (right
\label{fig:filament}
\end{figure}

To prove this hypothesis, first, a separate experiment in a 1-inch tube (same %gas flowrates and pressure 
reactor parameters as described in Section \ref{Setup}) was conducted to exaggerate the spatial dimension and better visualize any potential constricted region. The result is shown in Fig. \ref{fig:filament}(left), in which a bright straight columnar structure in the middle of the reactor can be seen. This observation is then converted %in 
into a model representation shown in Fig.\ref{fig:filament}(right). To connect the hot wire observation to the measurement results, the applied power portion absorbed by the plasma (i.e. electrons) and not the sheath must be determined. A formula derived by Godyak $et$ $al.$\cite{godyak1991} was used. Eq. \ref{eq:power_ratio} represents the ratio between the power absorbed by the ions in the sheath and the power absorbed by electrons:

\begin{equation}
\frac{P_i}{P_e}  = \frac{3 v_s}{2 d \nu_{en}} \bigg(\frac{\omega_e}{\omega}\bigg)^2
\label{eq:power_ratio}
\end{equation} 
Here, $v_s=\sqrt{\frac{eqT_e}{M}}$ is the ion Bohm velocity at the plasma-sheath edge, with $e$ being the electron charge, $q$ %is 
being the state of charge of ions, and $T_e$ being the electron temperature. Since CH$_4$ is a minor fraction of the feedstock mixture dominated by the argon flow, it is plausible to assume that the gas properties are largely defined by those of argon. Hence, $M$ can be approximated as the argon atom mass. With $T_e=2$ eV,\cite{godyak1990} $M=6.6\times10^{-26}$ kg, $v_s$ becomes $2.2\times10^3$  ms$^{-1}$. $\omega_e=\sqrt{\frac{e^2 n_e}{m_e \epsilon_0}}$ is the plasma frequency with $n_e$ being electron concentration $m_e$ being the electron mass and $\epsilon_0$ being the vacuum permittivity. With $n_e=10^{17}$ m$^{-3}$,\cite{moravej2004, godyak1990} $m_e=9.1\times10^{-31}$  kg, and $\epsilon_0=8.85\times10^{-12}$ F/m, $\omega_e$ becomes $1.78\times10^{10}$  s$^{-1}$. $\nu_{en}$ is the electron-neutral collision frequency. Using a scaling for argon from Raizer\cite{raizer1997} as $\nu_{en}=p\times5.3\times10^9$ Torr $\times$ s$^{-1}$Torr$^{-1}$, $\nu_{en}$ becomes $2.12\times10^{10}$ s$^{-1}$ for $p=4$ Torr. %With the reactor tube half width of $d=5$ mm 
With $d=L/2=1.75$ cm, where $L$ is the distance between the electrodes as shown in Fig. \ref{fig:schematic_reactor}, and the source frequency of 13.56 MHz converted to $\omega=8.52\times10^6$ s$^{-1}$, we finally find $P_i/P_e=53.87$. Next, the input power $P_{in}$ can be written as $P_{in}=P_i+P_e=P_e\bigg(1+\frac{P_i}{P_e}\bigg)$. Then, a new relation for the finalized $P_e$ can be written:

\begin{equation}
P_e  = \frac{\gamma P_{in}}{1+ \frac{P_i}{P_e}} 
\label{eq:power_electron}
\end{equation} 
where $\gamma$ is the matchbox efficiency. With assumed $\gamma$=0.7, the nominal input power of 200 W results in 8.5 W given to the electrons.

Because the ionization degree of the studied discharge was low, it is then plausible to treat the reactor medium as a neutral gas. Further on, assuming that the gas properties are largely defined by those of argon, the plasma is approximated as a thin hot wire source. Mathematically, the hot wire model, shown in Fig. \ref{fig:filament}(right), can be expressed by the gas heating $\Delta T$\cite{assael2023} as:

\begin{equation}
    \Delta T(r, t)  = \frac{P_e/L}{4 \pi \lambda} ln \bigg(\frac{4 \alpha t}{r^2 C}\bigg) 
\label{eq:gas_heating}
\end{equation}
where $C=e^{0.577}=1.781$, $\alpha$ is the thermal diffusivity, $\lambda$ is the thermal conductivity, and %$L$ is the plasma length such that 
$P_e⁄L$ is the linear power density. Eq. \ref{eq:gas_heating} quantifies the gas heating $\Delta T$ relative to the room temperature of 300 K at different coordinates $r$, from the reactor center to the wall, as a function of time $t$. 

\begin{figure*}[t]
\centering
    \includegraphics[width=\textwidth]{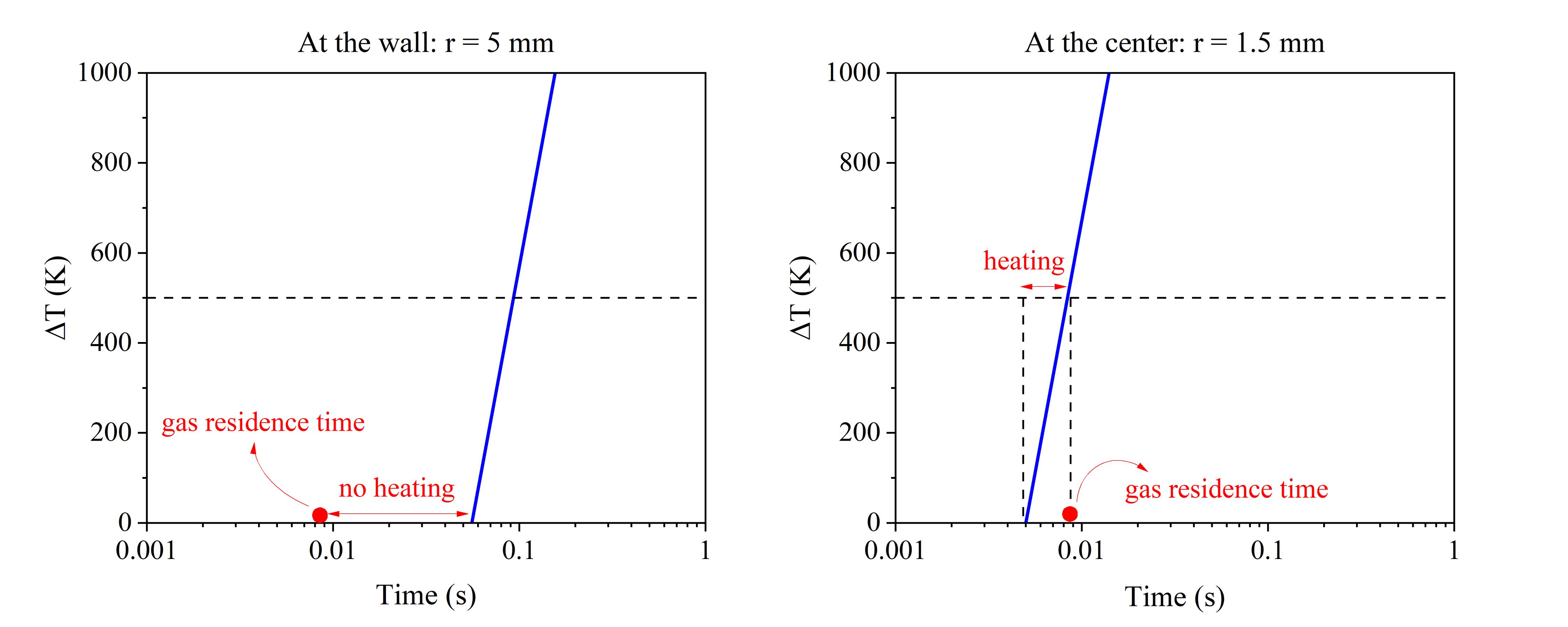}
    \caption{The temporal dependence of the relative temperature increase according to Eq. \ref{eq:gas_heating} with the specified parameters. These calculations do not include the heating taking place at the metal electrodes.}
\label{fig:temporal_heating}
\end{figure*}

Since the atom-atom mean free path is much smaller than the radius of the tube, we use the thermophysical data presented in Ref. \onlinecite{sun2005} to find $\alpha=20\times10^{-5}$ m$^2$s$^{-1}$ and $\lambda=0.02$ WmK$^{-1}$. From Eq. \ref{eq:gas_heating}, the plot presented in Fig. \ref{fig:temporal_heating} (left), shows a striking result: to enhance a measurable temperature increase on the wall, the gas flowing through the reactor has to spend at least $\sim0.06$ s. At the same time, the residence time of the gas $\tau$ is only 0.0087 s, calculated as follows:

\begin{equation}
    \tau  = \frac{\pi d^2 L}{\dot{q} \frac{p_a}{p}} \frac{60}{1}
\label{eq:residence_time}
\end{equation}
where $\dot{q}$ is the flow rate in %standard cubic centimeters per minute 
sccm, $p_a$ is the atmospheric pressure of 760 Torr and $p$ is the operating reactor pressure, and 60/1 is the minute-to-second conversion. With $p=4$ Torr, Eq. \ref{eq:residence_time} yields the noted residence time of 0.0087 %seconds 
s and a gas flow velocity of 4.03 %meters per second
ms$^{-1}$. On the other hand, at the center of tube, as shown in Fig. \ref{fig:temporal_heating} (right), this residence time is enough to increase the gas temperature by $\sim500$ K above room temperature. The analytically obtained distance of 1.5 mm of gas heating from the center of the tube appears to be consistent with the filament dimension observed visually in the reactor. 

Hence, from the model results it is seen that the main body (right outside the hot filament) of the discharge is cold. These results now can be used to explain the synthetis results. Once methyl radical is generated from a methane molecule, the activation energy for the initial step $^{\bullet}$CH$_3$ $+$ $^{\bullet}$CH$_3$ $\rightarrow$ $^{\bullet}$C$_2$H$_6$ to eventually form an adamantane molecule is high, close or above 13 kJmol$^{-1}$ with a reaction rate $<\sim10^{-11}$ cm$^3$s$^{-1}$.\cite{forst1991} To counter this limited kinetics, a higher gas temperature is required. At the same time, $sp^2$ graphitic reactions are expected to be virtually barrierless\cite{fedoseev1979} explaining immense product formation in cold Ar/CH$_4$ discharge. The cold gas medium does not allow $sp^2$ to crystallize leaving it in an amorphous state. While the same process should be happening in H$_2$/CH$_4$, $sp^2$ phase is immediately etched by atomic hydrogen with net zero product effect. The nonexistence of Si allotropy is the main reason of successful and well-developed Si syntheses with capacitively coupled RF continuous flow-through reactors, where $^{\bullet}$SiH$_3$ $+$ $^{\bullet}$SiH$_3$ reactions are barrierless.\cite{matsui1987}

\begin{figure}[h]
\centering
    \includegraphics[width=0.4\textwidth]{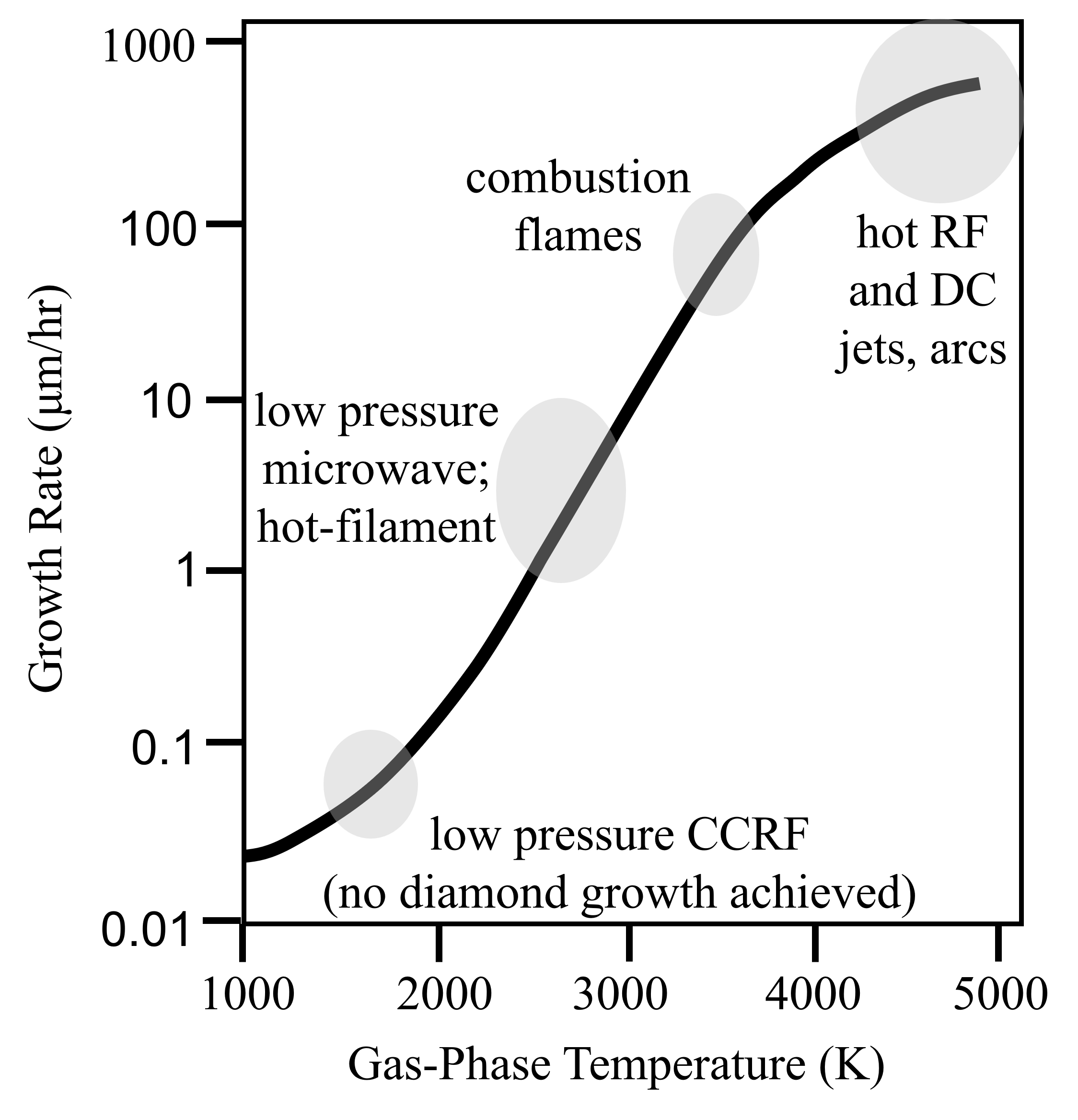}
    \caption{A comparison chart of the main diamond synthesis techniques with the addition of the low pressure RF plasma where no diamond growth has been achieved (adapted from Ref \onlinecite{bachmann1991}).}
\label{fig:comp_diam_synth_tech}
\end{figure}

The limiting nature of methyl radical chemistry in the gas phase or on the substrate\cite{kang2000} explains well why CVD reactors have to operate at high gas and substrate temperatures. From this, various diamond reactors can be compared as shown in Fig. \ref{fig:comp_diam_synth_tech}. Diamond growth starts at gas temperatures above 1,000 K. In the early 80s, small tubular DC flame, jet or arc reactors were successfully developed and used for diamond synthesis at high production rates.\cite{haubner2021} These designs were obviously successful because their operating gas temperatures are near 5,000 K allowing for fast diamond kinetics. At the same time, such designs were quickly ruled out due to scalability issues and cavity microwave reactors became the industrial workhorses for single crystal diamond wafer productions. 

In the new era of 3D additive manufacturing, reactor compactness is a plus, and flames, jets and arcs could be reconsidered for nanodiamond production. Talking about flow-through reactors, a number of new reactor strategies can be proposed. For CCRF, ambient pressure reactors of very small diameters could be beneficial for nanodiamond production. The plasma constriction (aka hot filament) is an unstable effect, therefore the easiest way to seize the hot gas filament is to induce it in a narrow capillary. High pressure increases RF power to electron coupling because, according to Eq. \ref{eq:power_ratio}, $P_i/P_e$ ratio decreases. For larger reactor volumes, inductively coupled flow-through systems could be useful such that gas heating is optimized for best diamond chemistry/kinetics. Third option could be using microwave flow through reactors where gas temperature can be manipulated by the same pressure-power product but in much wider ranges over a wider range of reactor volumes thanks to more efficient microwave to plasma power coupling.

\begin{comment}

\begin{figure}    
    \begin{subfigure}{0.23\textwidth}
       \includegraphics[width=\linewidth]{S3_5.png}     
     \label{fig:NCD_on_SCD_SEM1}   
    \end{subfigure}
    \begin{subfigure}{0.23\textwidth}
       \includegraphics[width=\linewidth]{S1_1.png}     
     \label{fig:NCD_on_SCD_SEM2}   
    \end{subfigure}

     \begin{subfigure}{0.5\textwidth}
        \includegraphics[width=\linewidth]{NCD_on_SCD_Raman.png}      
     \label{fig:NCD_on_SCD_Raman}
    \end{subfigure}
\caption{(top) SEM images, and (bottom) Raman spectrum of nanodiamond film deposited on SCD substrate via in-plasma collection.}
\label{fig:NCD_on_SCD}
\end{figure}

\end{comment}

\section{\label{sec:Conclusion}Conclusion}
A flow-through CCRF reactor utilizing Ar/CH$_4$ mixture was attempted to synthesize diamond nanocrystals. For standard conditions 4 Torr and 200 W, typical for diamond-cubic Si and Ge nanocrystal syntheses, final product material analyses revealed that only a $sp^2$ graphitic nanomaterial could be obtained, and with a fraction of $sp^3$ diamond nanocrystals (if any at all) that could not be registered experimentally. Using high-resolution LIF and OES diagnostics of the reactor, it was shown that power density and hence gas heating is concentrated in the narrow 1.5 mm on the axial region of the reactor, while the remainder of the plasma volume contains cold gas. This promotes the formation of the thermodynamically favorable graphitic allotrope to form, although some small amount of $sp^3$ hybridization in the material cannot be ruled out. If the same reactor was switched to operate with H$_2$/CH$_4$ mixture (traditionally used for diamond synthesis), no product was detected to form because atomic hydrogen actively must have etched the $sp^2$ hybridized nanomaterial, thereby leading to a net zero synthesis effect. These results put into context the legacy polycrystalline diamond synthetic routes, and outline practical strategies that could allow for future diamond nanocrystal feedstock production using flow-through low-temperature plasma reactors for bottom up additive manufacturing.

\section*{Acknowledgement}
This work was supported by the U.S. Department of Energy, Office of Science, Office of Fusion Energy Sciences, Award No. DE-SC0023211, and National Science Foundation, Division of Chemical, Bioengineering, Environmental and Transport Systems, Award No. 2333452. Work at Princeton Collaborative Low Temperature Plasma Research Facility (PCRF) was supported by the U.S. Department of Energy under Contract No. DE-AC02-09CH11466.

\section*{Data Access}
The data that support the findings of this study are available from the corresponding author upon reasonable request.

\section*{Conflict of Interest}
The authors have no conflicts to disclose.

\bibliography{Manuscript}% Produces the bibliography via BibTeX.

\begin{thebibliography}{57}%
\makeatletter
\providecommand \@ifxundefined [1]{%
 \@ifx{#1\undefined}
}%
\providecommand \@ifnum [1]{%
 \ifnum #1\expandafter \@firstoftwo
 \else \expandafter \@secondoftwo
 \fi
}%
\providecommand \@ifx [1]{%
 \ifx #1\expandafter \@firstoftwo
 \else \expandafter \@secondoftwo
 \fi
}%
\providecommand \natexlab [1]{#1}%
\providecommand \enquote  [1]{``#1''}%
\providecommand \bibnamefont  [1]{#1}%
\providecommand \bibfnamefont [1]{#1}%
\providecommand \citenamefont [1]{#1}%
\providecommand \href@noop [0]{\@secondoftwo}%
\providecommand \href [0]{\begingroup \@sanitize@url \@href}%
\providecommand \@href[1]{\@@startlink{#1}\@@href}%
\providecommand \@@href[1]{\endgroup#1\@@endlink}%
\providecommand \@sanitize@url [0]{\catcode `\\12\catcode `\$12\catcode `\&12\catcode `\#12\catcode `\^12\catcode `\_12\catcode `\%12\relax}%
\providecommand \@@startlink[1]{}%
\providecommand \@@endlink[0]{}%
\providecommand \url  [0]{\begingroup\@sanitize@url \@url }%
\providecommand \@url [1]{\endgroup\@href {#1}{\urlprefix }}%
\providecommand \urlprefix  [0]{URL }%
\providecommand \Eprint [0]{\href }%
\providecommand \doibase [0]{http://dx.doi.org/}%
\providecommand \selectlanguage [0]{\@gobble}%
\providecommand \bibinfo  [0]{\@secondoftwo}%
\providecommand \bibfield  [0]{\@secondoftwo}%
\providecommand \translation [1]{[#1]}%
\providecommand \BibitemOpen [0]{}%
\providecommand \bibitemStop [0]{}%
\providecommand \bibitemNoStop [0]{.\EOS\space}%
\providecommand \EOS [0]{\spacefactor3000\relax}%
\providecommand \BibitemShut  [1]{\csname bibitem#1\endcsname}%
\let\auto@bib@innerbib\@empty
%</preamble>
\bibitem [{\citenamefont {Graves}\ \emph {et~al.}(2023)\citenamefont {Graves}, \citenamefont {Labelle},\ and\ \citenamefont {Kushner}}]{graves2023}%
  \BibitemOpen
  \bibfield  {author} {\bibinfo {author} {\bibfnamefont {D.}~\bibnamefont {Graves}}, \bibinfo {author} {\bibfnamefont {C.}~\bibnamefont {Labelle}}, \ and\ \bibinfo {author} {\bibfnamefont {M.}~\bibnamefont {Kushner}},\ }\href@noop {} {\bibfield  {journal} {\bibinfo  {journal} {Department of Energy Office of Science, Fusion Energy Sciences}\ } (\bibinfo {year} {2023})}\BibitemShut {NoStop}%
\bibitem [{\citenamefont {Hori}\ and\ \citenamefont {Niemira}(2017)}]{hori2017}%
  \BibitemOpen
  \bibfield  {author} {\bibinfo {author} {\bibfnamefont {M.}~\bibnamefont {Hori}}\ and\ \bibinfo {author} {\bibfnamefont {B.}~\bibnamefont {Niemira}},\ }\href@noop {} {\bibfield  {journal} {\bibinfo  {journal} {J. Phys. D: Appl. Phys}\ }\textbf {\bibinfo {volume} {50}},\ \bibinfo {pages} {323001} (\bibinfo {year} {2017})}\BibitemShut {NoStop}%
\bibitem [{\citenamefont {Kortshagen}\ \emph {et~al.}(2016)\citenamefont {Kortshagen}, \citenamefont {Sankaran}, \citenamefont {Pereira}, \citenamefont {Girshick}, \citenamefont {Wu},\ and\ \citenamefont {Aydil}}]{kortshagen2016}%
  \BibitemOpen
  \bibfield  {author} {\bibinfo {author} {\bibfnamefont {U.~R.}\ \bibnamefont {Kortshagen}}, \bibinfo {author} {\bibfnamefont {R.~M.}\ \bibnamefont {Sankaran}}, \bibinfo {author} {\bibfnamefont {R.~N.}\ \bibnamefont {Pereira}}, \bibinfo {author} {\bibfnamefont {S.~L.}\ \bibnamefont {Girshick}}, \bibinfo {author} {\bibfnamefont {J.~J.}\ \bibnamefont {Wu}}, \ and\ \bibinfo {author} {\bibfnamefont {E.~S.}\ \bibnamefont {Aydil}},\ }\href@noop {} {\bibfield  {journal} {\bibinfo  {journal} {Chemical reviews}\ }\textbf {\bibinfo {volume} {116}},\ \bibinfo {pages} {11061} (\bibinfo {year} {2016})}\BibitemShut {NoStop}%
\bibitem [{\citenamefont {Anthony}\ and\ \citenamefont {Kortshagen}(2009)}]{anthony2009}%
  \BibitemOpen
  \bibfield  {author} {\bibinfo {author} {\bibfnamefont {R.}~\bibnamefont {Anthony}}\ and\ \bibinfo {author} {\bibfnamefont {U.}~\bibnamefont {Kortshagen}},\ }\href@noop {} {\bibfield  {journal} {\bibinfo  {journal} {Phys. Rev. B}\ }\textbf {\bibinfo {volume} {80}},\ \bibinfo {pages} {115407} (\bibinfo {year} {2009})}\BibitemShut {NoStop}%
\bibitem [{\citenamefont {Kortshagen}(2016)}]{kortshagen2016_springer}%
  \BibitemOpen
  \bibfield  {author} {\bibinfo {author} {\bibfnamefont {U.}~\bibnamefont {Kortshagen}},\ }\href@noop {} {\bibfield  {journal} {\bibinfo  {journal} {Plasma Chemistry and Plasma Processing}\ }\textbf {\bibinfo {volume} {36}},\ \bibinfo {pages} {73} (\bibinfo {year} {2016})}\BibitemShut {NoStop}%
\bibitem [{\citenamefont {Mangolini}\ and\ \citenamefont {Kortshagen}(2007)}]{mangolini2007}%
  \BibitemOpen
  \bibfield  {author} {\bibinfo {author} {\bibfnamefont {L.}~\bibnamefont {Mangolini}}\ and\ \bibinfo {author} {\bibfnamefont {U.}~\bibnamefont {Kortshagen}},\ }\href@noop {} {\bibfield  {journal} {\bibinfo  {journal} {Advanced Materials}\ }\textbf {\bibinfo {volume} {19}} (\bibinfo {year} {2007})}\BibitemShut {NoStop}%
\bibitem [{\citenamefont {Woodard}\ \emph {et~al.}(2018)\citenamefont {Woodard}, \citenamefont {Xu}, \citenamefont {Barragan}, \citenamefont {Nava}, \citenamefont {Wong},\ and\ \citenamefont {Mangolini}}]{woodard2018}%
  \BibitemOpen
  \bibfield  {author} {\bibinfo {author} {\bibfnamefont {A.}~\bibnamefont {Woodard}}, \bibinfo {author} {\bibfnamefont {L.}~\bibnamefont {Xu}}, \bibinfo {author} {\bibfnamefont {A.~A.}\ \bibnamefont {Barragan}}, \bibinfo {author} {\bibfnamefont {G.}~\bibnamefont {Nava}}, \bibinfo {author} {\bibfnamefont {B.~M.}\ \bibnamefont {Wong}}, \ and\ \bibinfo {author} {\bibfnamefont {L.}~\bibnamefont {Mangolini}},\ }\href@noop {} {\bibfield  {journal} {\bibinfo  {journal} {Plasma Processes and Polymers}\ }\textbf {\bibinfo {volume} {15}},\ \bibinfo {pages} {1700104} (\bibinfo {year} {2018})}\BibitemShut {NoStop}%
\bibitem [{\citenamefont {Exarhos}\ \emph {et~al.}(2018)\citenamefont {Exarhos}, \citenamefont {Alvarez-Barragan}, \citenamefont {Aytan}, \citenamefont {Balandin},\ and\ \citenamefont {Mangolini}}]{exarhos2018}%
  \BibitemOpen
  \bibfield  {author} {\bibinfo {author} {\bibfnamefont {S.}~\bibnamefont {Exarhos}}, \bibinfo {author} {\bibfnamefont {A.}~\bibnamefont {Alvarez-Barragan}}, \bibinfo {author} {\bibfnamefont {E.}~\bibnamefont {Aytan}}, \bibinfo {author} {\bibfnamefont {A.~A.}\ \bibnamefont {Balandin}}, \ and\ \bibinfo {author} {\bibfnamefont {L.}~\bibnamefont {Mangolini}},\ }\href@noop {} {\bibfield  {journal} {\bibinfo  {journal} {ACS Energy Letters}\ }\textbf {\bibinfo {volume} {3}},\ \bibinfo {pages} {2349} (\bibinfo {year} {2018})}\BibitemShut {NoStop}%
\bibitem [{\citenamefont {Thimsen}\ \emph {et~al.}(2015)\citenamefont {Thimsen}, \citenamefont {Kortshagen},\ and\ \citenamefont {Aydil}}]{thimsen2015}%
  \BibitemOpen
  \bibfield  {author} {\bibinfo {author} {\bibfnamefont {E.}~\bibnamefont {Thimsen}}, \bibinfo {author} {\bibfnamefont {U.~R.}\ \bibnamefont {Kortshagen}}, \ and\ \bibinfo {author} {\bibfnamefont {E.~S.}\ \bibnamefont {Aydil}},\ }\href@noop {} {\bibfield  {journal} {\bibinfo  {journal} {Journal of Physics D: Applied Physics}\ }\textbf {\bibinfo {volume} {48}},\ \bibinfo {pages} {314004} (\bibinfo {year} {2015})}\BibitemShut {NoStop}%
\bibitem [{\citenamefont {Uner}\ \emph {et~al.}(2019)\citenamefont {Uner}, \citenamefont {Niedzwiedzki},\ and\ \citenamefont {Thimsen}}]{uner2019}%
  \BibitemOpen
  \bibfield  {author} {\bibinfo {author} {\bibfnamefont {N.~B.}\ \bibnamefont {Uner}}, \bibinfo {author} {\bibfnamefont {D.~M.}\ \bibnamefont {Niedzwiedzki}}, \ and\ \bibinfo {author} {\bibfnamefont {E.}~\bibnamefont {Thimsen}},\ }\href@noop {} {\bibfield  {journal} {\bibinfo  {journal} {The Journal of Physical Chemistry C}\ }\textbf {\bibinfo {volume} {123}},\ \bibinfo {pages} {30613} (\bibinfo {year} {2019})}\BibitemShut {NoStop}%
\bibitem [{\citenamefont {Izadi}\ and\ \citenamefont {Anthony}(2019)}]{izadi2019}%
  \BibitemOpen
  \bibfield  {author} {\bibinfo {author} {\bibfnamefont {A.}~\bibnamefont {Izadi}}\ and\ \bibinfo {author} {\bibfnamefont {R.~J.}\ \bibnamefont {Anthony}},\ }\href@noop {} {\bibfield  {journal} {\bibinfo  {journal} {Plasma Processes and Polymers}\ }\textbf {\bibinfo {volume} {16}},\ \bibinfo {pages} {e1800212} (\bibinfo {year} {2019})}\BibitemShut {NoStop}%
\bibitem [{\citenamefont {Mandal}\ \emph {et~al.}(2018)\citenamefont {Mandal}, \citenamefont {O’Shea},\ and\ \citenamefont {Anthony}}]{mandal2018}%
  \BibitemOpen
  \bibfield  {author} {\bibinfo {author} {\bibfnamefont {R.}~\bibnamefont {Mandal}}, \bibinfo {author} {\bibfnamefont {K.}~\bibnamefont {O’Shea}}, \ and\ \bibinfo {author} {\bibfnamefont {R.}~\bibnamefont {Anthony}},\ }\href@noop {} {\bibfield  {journal} {\bibinfo  {journal} {Journal of Vacuum Science \& Technology A}\ }\textbf {\bibinfo {volume} {36}} (\bibinfo {year} {2018})}\BibitemShut {NoStop}%
\bibitem [{\citenamefont {Ho}\ \emph {et~al.}(2021)\citenamefont {Ho}, \citenamefont {Mandal}, \citenamefont {Lunt},\ and\ \citenamefont {Anthony}}]{ho2021}%
  \BibitemOpen
  \bibfield  {author} {\bibinfo {author} {\bibfnamefont {A.}~\bibnamefont {Ho}}, \bibinfo {author} {\bibfnamefont {R.}~\bibnamefont {Mandal}}, \bibinfo {author} {\bibfnamefont {R.~R.}\ \bibnamefont {Lunt}}, \ and\ \bibinfo {author} {\bibfnamefont {R.~J.}\ \bibnamefont {Anthony}},\ }\href@noop {} {\bibfield  {journal} {\bibinfo  {journal} {ACS Applied Nano Materials}\ }\textbf {\bibinfo {volume} {4}},\ \bibinfo {pages} {5624} (\bibinfo {year} {2021})}\BibitemShut {NoStop}%
\bibitem [{\citenamefont {Dsouza}\ \emph {et~al.}(2023)\citenamefont {Dsouza}, \citenamefont {Buerkle}, \citenamefont {Alessi}, \citenamefont {Brunet}, \citenamefont {Morelli}, \citenamefont {Payam}, \citenamefont {Maguire}, \citenamefont {Mariotti},\ and\ \citenamefont {Svrcek}}]{dsouza2023}%
  \BibitemOpen
  \bibfield  {author} {\bibinfo {author} {\bibfnamefont {S.~D.}\ \bibnamefont {Dsouza}}, \bibinfo {author} {\bibfnamefont {M.}~\bibnamefont {Buerkle}}, \bibinfo {author} {\bibfnamefont {B.}~\bibnamefont {Alessi}}, \bibinfo {author} {\bibfnamefont {P.}~\bibnamefont {Brunet}}, \bibinfo {author} {\bibfnamefont {A.}~\bibnamefont {Morelli}}, \bibinfo {author} {\bibfnamefont {A.~F.}\ \bibnamefont {Payam}}, \bibinfo {author} {\bibfnamefont {P.}~\bibnamefont {Maguire}}, \bibinfo {author} {\bibfnamefont {D.}~\bibnamefont {Mariotti}}, \ and\ \bibinfo {author} {\bibfnamefont {V.}~\bibnamefont {Svrcek}},\ }\href@noop {} {\bibfield  {journal} {\bibinfo  {journal} {Nanotechnology}\ }\textbf {\bibinfo {volume} {34}},\ \bibinfo {pages} {505601} (\bibinfo {year} {2023})}\BibitemShut {NoStop}%
\bibitem [{\citenamefont {Askari}\ \emph {et~al.}(2015)\citenamefont {Askari}, \citenamefont {Macias-Montero}, \citenamefont {Velusamy}, \citenamefont {Maguire}, \citenamefont {Svrcek},\ and\ \citenamefont {Mariotti}}]{askari2015}%
  \BibitemOpen
  \bibfield  {author} {\bibinfo {author} {\bibfnamefont {S.}~\bibnamefont {Askari}}, \bibinfo {author} {\bibfnamefont {M.}~\bibnamefont {Macias-Montero}}, \bibinfo {author} {\bibfnamefont {T.}~\bibnamefont {Velusamy}}, \bibinfo {author} {\bibfnamefont {P.}~\bibnamefont {Maguire}}, \bibinfo {author} {\bibfnamefont {V.}~\bibnamefont {Svrcek}}, \ and\ \bibinfo {author} {\bibfnamefont {D.}~\bibnamefont {Mariotti}},\ }\href@noop {} {\bibfield  {journal} {\bibinfo  {journal} {Journal of Physics D: Applied Physics}\ }\textbf {\bibinfo {volume} {48}},\ \bibinfo {pages} {314002} (\bibinfo {year} {2015})}\BibitemShut {NoStop}%
\bibitem [{\citenamefont {Nozaki}\ \emph {et~al.}(2007)\citenamefont {Nozaki}, \citenamefont {Sasaki}, \citenamefont {Ogino}, \citenamefont {Asahi},\ and\ \citenamefont {Okazaki}}]{nozaki2007}%
  \BibitemOpen
  \bibfield  {author} {\bibinfo {author} {\bibfnamefont {T.}~\bibnamefont {Nozaki}}, \bibinfo {author} {\bibfnamefont {K.}~\bibnamefont {Sasaki}}, \bibinfo {author} {\bibfnamefont {T.}~\bibnamefont {Ogino}}, \bibinfo {author} {\bibfnamefont {D.}~\bibnamefont {Asahi}}, \ and\ \bibinfo {author} {\bibfnamefont {K.}~\bibnamefont {Okazaki}},\ }\href@noop {} {\bibfield  {journal} {\bibinfo  {journal} {Journal of Thermal Science and Technology}\ }\textbf {\bibinfo {volume} {2}},\ \bibinfo {pages} {192} (\bibinfo {year} {2007})}\BibitemShut {NoStop}%
\bibitem [{\citenamefont {Mangolini}\ \emph {et~al.}(2005)\citenamefont {Mangolini}, \citenamefont {Thimsen},\ and\ \citenamefont {Kortshagen}}]{mangolini2005}%
  \BibitemOpen
  \bibfield  {author} {\bibinfo {author} {\bibfnamefont {L.}~\bibnamefont {Mangolini}}, \bibinfo {author} {\bibfnamefont {E.}~\bibnamefont {Thimsen}}, \ and\ \bibinfo {author} {\bibfnamefont {U.}~\bibnamefont {Kortshagen}},\ }\href@noop {} {\bibfield  {journal} {\bibinfo  {journal} {Nano letters}\ }\textbf {\bibinfo {volume} {5}},\ \bibinfo {pages} {655} (\bibinfo {year} {2005})}\BibitemShut {NoStop}%
\bibitem [{\citenamefont {Pi}\ and\ \citenamefont {Kortshagen}(2009)}]{pi2009}%
  \BibitemOpen
  \bibfield  {author} {\bibinfo {author} {\bibfnamefont {X.}~\bibnamefont {Pi}}\ and\ \bibinfo {author} {\bibfnamefont {U.}~\bibnamefont {Kortshagen}},\ }\href@noop {} {\bibfield  {journal} {\bibinfo  {journal} {Nanotechnology}\ }\textbf {\bibinfo {volume} {20}},\ \bibinfo {pages} {295602} (\bibinfo {year} {2009})}\BibitemShut {NoStop}%
\bibitem [{\citenamefont {Gresback}\ \emph {et~al.}(2007)\citenamefont {Gresback}, \citenamefont {Holman},\ and\ \citenamefont {Kortshagen}}]{gresback2007}%
  \BibitemOpen
  \bibfield  {author} {\bibinfo {author} {\bibfnamefont {R.}~\bibnamefont {Gresback}}, \bibinfo {author} {\bibfnamefont {Z.}~\bibnamefont {Holman}}, \ and\ \bibinfo {author} {\bibfnamefont {U.}~\bibnamefont {Kortshagen}},\ }\href@noop {} {\bibfield  {journal} {\bibinfo  {journal} {Applied Physics Letters}\ }\textbf {\bibinfo {volume} {91}} (\bibinfo {year} {2007})}\BibitemShut {NoStop}%
\bibitem [{\citenamefont {Deryagin}\ and\ \citenamefont {Fedoseev}(1979)}]{deryagin1979}%
  \BibitemOpen
  \bibfield  {author} {\bibinfo {author} {\bibfnamefont {B.~V.}\ \bibnamefont {Deryagin}}\ and\ \bibinfo {author} {\bibfnamefont {D.~V.}\ \bibnamefont {Fedoseev}},\ }\href@noop {} {\bibfield  {journal} {\bibinfo  {journal} {Bulletin of the Academy of Sciences of the USSR, Division of chemical science}\ }\textbf {\bibinfo {volume} {28}},\ \bibinfo {pages} {1106} (\bibinfo {year} {1979})}\BibitemShut {NoStop}%
\bibitem [{\citenamefont {Frenklach}\ \emph {et~al.}(1989)\citenamefont {Frenklach}, \citenamefont {Kematick}, \citenamefont {Huang}, \citenamefont {Howard}, \citenamefont {Spear}, \citenamefont {Phelps},\ and\ \citenamefont {Koba}}]{frenklach1989}%
  \BibitemOpen
  \bibfield  {author} {\bibinfo {author} {\bibfnamefont {M.}~\bibnamefont {Frenklach}}, \bibinfo {author} {\bibfnamefont {R.}~\bibnamefont {Kematick}}, \bibinfo {author} {\bibfnamefont {D.}~\bibnamefont {Huang}}, \bibinfo {author} {\bibfnamefont {W.}~\bibnamefont {Howard}}, \bibinfo {author} {\bibfnamefont {K.}~\bibnamefont {Spear}}, \bibinfo {author} {\bibfnamefont {A.~W.}\ \bibnamefont {Phelps}}, \ and\ \bibinfo {author} {\bibfnamefont {R.}~\bibnamefont {Koba}},\ }\href@noop {} {\bibfield  {journal} {\bibinfo  {journal} {Journal of applied physics}\ }\textbf {\bibinfo {volume} {66}},\ \bibinfo {pages} {395} (\bibinfo {year} {1989})}\BibitemShut {NoStop}%
\bibitem [{\citenamefont {Kumar}\ \emph {et~al.}(2013)\citenamefont {Kumar}, \citenamefont {Ann~Lin}, \citenamefont {Xue}, \citenamefont {Hao}, \citenamefont {Khin~Yap},\ and\ \citenamefont {Sankaran}}]{kumar2013}%
  \BibitemOpen
  \bibfield  {author} {\bibinfo {author} {\bibfnamefont {A.}~\bibnamefont {Kumar}}, \bibinfo {author} {\bibfnamefont {P.}~\bibnamefont {Ann~Lin}}, \bibinfo {author} {\bibfnamefont {A.}~\bibnamefont {Xue}}, \bibinfo {author} {\bibfnamefont {B.}~\bibnamefont {Hao}}, \bibinfo {author} {\bibfnamefont {Y.}~\bibnamefont {Khin~Yap}}, \ and\ \bibinfo {author} {\bibfnamefont {R.~M.}\ \bibnamefont {Sankaran}},\ }\href@noop {} {\bibfield  {journal} {\bibinfo  {journal} {Nature communications}\ }\textbf {\bibinfo {volume} {4}},\ \bibinfo {pages} {2618} (\bibinfo {year} {2013})}\BibitemShut {NoStop}%
\bibitem [{\citenamefont {Auciello}(2022)}]{auciello2022}%
  \BibitemOpen
  \bibfield  {author} {\bibinfo {author} {\bibfnamefont {O.}~\bibnamefont {Auciello}},\ }\href@noop {} {\bibfield  {journal} {\bibinfo  {journal} {Functional Diamond}\ }\textbf {\bibinfo {volume} {2}},\ \bibinfo {pages} {1} (\bibinfo {year} {2022})}\BibitemShut {NoStop}%
\bibitem [{\citenamefont {Yatom}\ \emph {et~al.}(2018)\citenamefont {Yatom}, \citenamefont {Khrabry}, \citenamefont {Mitrani}, \citenamefont {Khodak}, \citenamefont {Kaganovich}, \citenamefont {Vekselman}, \citenamefont {Stratton},\ and\ \citenamefont {Raitses}}]{yatom2018_MRS}%
  \BibitemOpen
  \bibfield  {author} {\bibinfo {author} {\bibfnamefont {S.}~\bibnamefont {Yatom}}, \bibinfo {author} {\bibfnamefont {A.}~\bibnamefont {Khrabry}}, \bibinfo {author} {\bibfnamefont {J.}~\bibnamefont {Mitrani}}, \bibinfo {author} {\bibfnamefont {A.}~\bibnamefont {Khodak}}, \bibinfo {author} {\bibfnamefont {I.}~\bibnamefont {Kaganovich}}, \bibinfo {author} {\bibfnamefont {V.}~\bibnamefont {Vekselman}}, \bibinfo {author} {\bibfnamefont {B.}~\bibnamefont {Stratton}}, \ and\ \bibinfo {author} {\bibfnamefont {Y.}~\bibnamefont {Raitses}},\ }\href@noop {} {\bibfield  {journal} {\bibinfo  {journal} {MRS communications}\ }\textbf {\bibinfo {volume} {8}},\ \bibinfo {pages} {842} (\bibinfo {year} {2018})}\BibitemShut {NoStop}%
\bibitem [{\citenamefont {Yatom}\ \emph {et~al.}(2017)\citenamefont {Yatom}, \citenamefont {Bak}, \citenamefont {Khrabryi},\ and\ \citenamefont {Raitses}}]{yatom2017_carbon}%
  \BibitemOpen
  \bibfield  {author} {\bibinfo {author} {\bibfnamefont {S.}~\bibnamefont {Yatom}}, \bibinfo {author} {\bibfnamefont {J.}~\bibnamefont {Bak}}, \bibinfo {author} {\bibfnamefont {A.}~\bibnamefont {Khrabryi}}, \ and\ \bibinfo {author} {\bibfnamefont {Y.}~\bibnamefont {Raitses}},\ }\href@noop {} {\bibfield  {journal} {\bibinfo  {journal} {Carbon}\ }\textbf {\bibinfo {volume} {117}},\ \bibinfo {pages} {154} (\bibinfo {year} {2017})}\BibitemShut {NoStop}%
\bibitem [{\citenamefont {Stratton}\ \emph {et~al.}(2018)\citenamefont {Stratton}, \citenamefont {Gerakis}, \citenamefont {Kaganovich}, \citenamefont {Keidar}, \citenamefont {Khrabry}, \citenamefont {Mitrani}, \citenamefont {Raitses}, \citenamefont {Shneider}, \citenamefont {Vekselman},\ and\ \citenamefont {Yatom}}]{stratton2018}%
  \BibitemOpen
  \bibfield  {author} {\bibinfo {author} {\bibfnamefont {B.}~\bibnamefont {Stratton}}, \bibinfo {author} {\bibfnamefont {A.}~\bibnamefont {Gerakis}}, \bibinfo {author} {\bibfnamefont {I.}~\bibnamefont {Kaganovich}}, \bibinfo {author} {\bibfnamefont {M.}~\bibnamefont {Keidar}}, \bibinfo {author} {\bibfnamefont {A.}~\bibnamefont {Khrabry}}, \bibinfo {author} {\bibfnamefont {J.}~\bibnamefont {Mitrani}}, \bibinfo {author} {\bibfnamefont {Y.}~\bibnamefont {Raitses}}, \bibinfo {author} {\bibfnamefont {M.}~\bibnamefont {Shneider}}, \bibinfo {author} {\bibfnamefont {V.}~\bibnamefont {Vekselman}}, \ and\ \bibinfo {author} {\bibfnamefont {S.}~\bibnamefont {Yatom}},\ }\href@noop {} {\bibfield  {journal} {\bibinfo  {journal} {Plasma Sources Science and Technology}\ }\textbf {\bibinfo {volume} {27}},\ \bibinfo {pages} {084001} (\bibinfo {year} {2018})}\BibitemShut {NoStop}%
\bibitem [{\citenamefont {Vekselman}\ \emph {et~al.}(2018)\citenamefont {Vekselman}, \citenamefont {Khrabry}, \citenamefont {Kaganovich}, \citenamefont {Stratton}, \citenamefont {Selinsky},\ and\ \citenamefont {Raitses}}]{vekselman2018}%
  \BibitemOpen
  \bibfield  {author} {\bibinfo {author} {\bibfnamefont {V.}~\bibnamefont {Vekselman}}, \bibinfo {author} {\bibfnamefont {A.}~\bibnamefont {Khrabry}}, \bibinfo {author} {\bibfnamefont {I.}~\bibnamefont {Kaganovich}}, \bibinfo {author} {\bibfnamefont {B.}~\bibnamefont {Stratton}}, \bibinfo {author} {\bibfnamefont {R.}~\bibnamefont {Selinsky}}, \ and\ \bibinfo {author} {\bibfnamefont {Y.}~\bibnamefont {Raitses}},\ }\href@noop {} {\bibfield  {journal} {\bibinfo  {journal} {Plasma Sources Science and Technology}\ }\textbf {\bibinfo {volume} {27}},\ \bibinfo {pages} {025008} (\bibinfo {year} {2018})}\BibitemShut {NoStop}%
\bibitem [{\citenamefont {Yatom}\ and\ \citenamefont {Dobrynin}(2022)}]{yatom2022_physD}%
  \BibitemOpen
  \bibfield  {author} {\bibinfo {author} {\bibfnamefont {S.}~\bibnamefont {Yatom}}\ and\ \bibinfo {author} {\bibfnamefont {D.}~\bibnamefont {Dobrynin}},\ }\href@noop {} {\bibfield  {journal} {\bibinfo  {journal} {Journal of Physics D: Applied Physics}\ }\textbf {\bibinfo {volume} {55}},\ \bibinfo {pages} {485203} (\bibinfo {year} {2022})}\BibitemShut {NoStop}%
\bibitem [{\citenamefont {Yatom}\ \emph {et~al.}(2023)\citenamefont {Yatom}, \citenamefont {Chopra}, \citenamefont {Kondeti}, \citenamefont {Petrova}, \citenamefont {Raitses}, \citenamefont {Boris}, \citenamefont {Johnson},\ and\ \citenamefont {Walton}}]{yatom2023_plasma}%
  \BibitemOpen
  \bibfield  {author} {\bibinfo {author} {\bibfnamefont {S.}~\bibnamefont {Yatom}}, \bibinfo {author} {\bibfnamefont {N.}~\bibnamefont {Chopra}}, \bibinfo {author} {\bibfnamefont {S.}~\bibnamefont {Kondeti}}, \bibinfo {author} {\bibfnamefont {T.~B.}\ \bibnamefont {Petrova}}, \bibinfo {author} {\bibfnamefont {Y.}~\bibnamefont {Raitses}}, \bibinfo {author} {\bibfnamefont {D.~R.}\ \bibnamefont {Boris}}, \bibinfo {author} {\bibfnamefont {M.~J.}\ \bibnamefont {Johnson}}, \ and\ \bibinfo {author} {\bibfnamefont {S.~G.}\ \bibnamefont {Walton}},\ }\href@noop {} {\bibfield  {journal} {\bibinfo  {journal} {Plasma Sources Science and Technology}\ }\textbf {\bibinfo {volume} {32}},\ \bibinfo {pages} {115005} (\bibinfo {year} {2023})}\BibitemShut {NoStop}%
\bibitem [{\citenamefont {Luque}\ \emph {et~al.}(1997{\natexlab{a}})\citenamefont {Luque}, \citenamefont {Juchmann},\ and\ \citenamefont {Jeffries}}]{luque1997}%
  \BibitemOpen
  \bibfield  {author} {\bibinfo {author} {\bibfnamefont {J.}~\bibnamefont {Luque}}, \bibinfo {author} {\bibfnamefont {W.}~\bibnamefont {Juchmann}}, \ and\ \bibinfo {author} {\bibfnamefont {J.}~\bibnamefont {Jeffries}},\ }\href@noop {} {\bibfield  {journal} {\bibinfo  {journal} {Journal of applied physics}\ }\textbf {\bibinfo {volume} {82}},\ \bibinfo {pages} {2072} (\bibinfo {year} {1997}{\natexlab{a}})}\BibitemShut {NoStop}%
\bibitem [{\citenamefont {Yatom}\ and\ \citenamefont {Raitses}(2020)}]{yatom2020_phychem}%
  \BibitemOpen
  \bibfield  {author} {\bibinfo {author} {\bibfnamefont {S.}~\bibnamefont {Yatom}}\ and\ \bibinfo {author} {\bibfnamefont {Y.}~\bibnamefont {Raitses}},\ }\href@noop {} {\bibfield  {journal} {\bibinfo  {journal} {Physical Chemistry Chemical Physics}\ }\textbf {\bibinfo {volume} {22}},\ \bibinfo {pages} {20837} (\bibinfo {year} {2020})}\BibitemShut {NoStop}%
\bibitem [{\citenamefont {Galli}\ and\ \citenamefont {Kortshagen}(2009)}]{galli2009}%
  \BibitemOpen
  \bibfield  {author} {\bibinfo {author} {\bibfnamefont {F.}~\bibnamefont {Galli}}\ and\ \bibinfo {author} {\bibfnamefont {U.~R.}\ \bibnamefont {Kortshagen}},\ }\href@noop {} {\bibfield  {journal} {\bibinfo  {journal} {IEEE Transactions on Plasma Science}\ }\textbf {\bibinfo {volume} {38}},\ \bibinfo {pages} {803} (\bibinfo {year} {2009})}\BibitemShut {NoStop}%
\bibitem [{\citenamefont {Mangolini}\ and\ \citenamefont {Kortshagen}(2009)}]{mangolini2009}%
  \BibitemOpen
  \bibfield  {author} {\bibinfo {author} {\bibfnamefont {L.}~\bibnamefont {Mangolini}}\ and\ \bibinfo {author} {\bibfnamefont {U.}~\bibnamefont {Kortshagen}},\ }\href@noop {} {\bibfield  {journal} {\bibinfo  {journal} {Physical review E}\ }\textbf {\bibinfo {volume} {79}},\ \bibinfo {pages} {026405} (\bibinfo {year} {2009})}\BibitemShut {NoStop}%
\bibitem [{\citenamefont {Lopez}\ and\ \citenamefont {Mangolini}(2014)}]{lopez2014}%
  \BibitemOpen
  \bibfield  {author} {\bibinfo {author} {\bibfnamefont {T.}~\bibnamefont {Lopez}}\ and\ \bibinfo {author} {\bibfnamefont {L.}~\bibnamefont {Mangolini}},\ }\href@noop {} {\bibfield  {journal} {\bibinfo  {journal} {Journal of Vacuum Science and Technology B}\ }\textbf {\bibinfo {volume} {32}},\ \bibinfo {pages} {061802} (\bibinfo {year} {2014})}\BibitemShut {NoStop}%
\bibitem [{\citenamefont {Mahoney}\ \emph {et~al.}(2019)\citenamefont {Mahoney}, \citenamefont {Rodriguez}, \citenamefont {Mushtaq}, \citenamefont {Truscott}, \citenamefont {Ashfold},\ and\ \citenamefont {Mankelevich}}]{mahoney2019}%
  \BibitemOpen
  \bibfield  {author} {\bibinfo {author} {\bibfnamefont {E.~J.}\ \bibnamefont {Mahoney}}, \bibinfo {author} {\bibfnamefont {B.~J.}\ \bibnamefont {Rodriguez}}, \bibinfo {author} {\bibfnamefont {S.}~\bibnamefont {Mushtaq}}, \bibinfo {author} {\bibfnamefont {B.~S.}\ \bibnamefont {Truscott}}, \bibinfo {author} {\bibfnamefont {M.~N.}\ \bibnamefont {Ashfold}}, \ and\ \bibinfo {author} {\bibfnamefont {Y.~A.}\ \bibnamefont {Mankelevich}},\ }\href@noop {} {\bibfield  {journal} {\bibinfo  {journal} {The Journal of Physical Chemistry A}\ }\textbf {\bibinfo {volume} {123}},\ \bibinfo {pages} {9966} (\bibinfo {year} {2019})}\BibitemShut {NoStop}%
\bibitem [{\citenamefont {Luque}\ \emph {et~al.}(1997{\natexlab{b}})\citenamefont {Luque}, \citenamefont {Juchmann},\ and\ \citenamefont {Jeffries}}]{luque1997_optica}%
  \BibitemOpen
  \bibfield  {author} {\bibinfo {author} {\bibfnamefont {J.}~\bibnamefont {Luque}}, \bibinfo {author} {\bibfnamefont {W.}~\bibnamefont {Juchmann}}, \ and\ \bibinfo {author} {\bibfnamefont {J.}~\bibnamefont {Jeffries}},\ }\href@noop {} {\bibfield  {journal} {\bibinfo  {journal} {Applied optics}\ }\textbf {\bibinfo {volume} {36}},\ \bibinfo {pages} {3261} (\bibinfo {year} {1997}{\natexlab{b}})}\BibitemShut {NoStop}%
\bibitem [{\citenamefont {Hollas}(2004)}]{hollas2004}%
  \BibitemOpen
  \bibfield  {author} {\bibinfo {author} {\bibfnamefont {J.~M.}\ \bibnamefont {Hollas}},\ }\href@noop {} {\emph {\bibinfo {title} {Modern Spectroscopy, 4th Edition}}}\ (\bibinfo  {publisher} {Wiley},\ \bibinfo {year} {2004})\BibitemShut {NoStop}%
\bibitem [{\citenamefont {Huber}\ and\ \citenamefont {Herzberg}(1979)}]{huber1979}%
  \BibitemOpen
  \bibfield  {author} {\bibinfo {author} {\bibfnamefont {K.~P.}\ \bibnamefont {Huber}}\ and\ \bibinfo {author} {\bibfnamefont {G.}~\bibnamefont {Herzberg}},\ }\href@noop {} {\emph {\bibinfo {title} {Molecular Spectra and Molecular Structure}}}\ (\bibinfo  {publisher} {Springer},\ \bibinfo {year} {1979})\BibitemShut {NoStop}%
\bibitem [{\citenamefont {Partridge}\ and\ \citenamefont {Laurendeau}(1995)}]{partridge1995}%
  \BibitemOpen
  \bibfield  {author} {\bibinfo {author} {\bibfnamefont {W.~P.}\ \bibnamefont {Partridge}}\ and\ \bibinfo {author} {\bibfnamefont {N.~M.}\ \bibnamefont {Laurendeau}},\ }\href@noop {} {\bibfield  {journal} {\bibinfo  {journal} {Applied optics}\ }\textbf {\bibinfo {volume} {34}},\ \bibinfo {pages} {2645} (\bibinfo {year} {1995})}\BibitemShut {NoStop}%
\bibitem [{\citenamefont {Reichardt}\ \emph {et~al.}(2003)\citenamefont {Reichardt}, \citenamefont {Baumgart},\ and\ \citenamefont {McGee}}]{reichardt2003}%
  \BibitemOpen
  \bibfield  {author} {\bibinfo {author} {\bibfnamefont {J.}~\bibnamefont {Reichardt}}, \bibinfo {author} {\bibfnamefont {R.}~\bibnamefont {Baumgart}}, \ and\ \bibinfo {author} {\bibfnamefont {T.~J.}\ \bibnamefont {McGee}},\ }\href@noop {} {\bibfield  {journal} {\bibinfo  {journal} {Applied optics}\ }\textbf {\bibinfo {volume} {42}},\ \bibinfo {pages} {4909} (\bibinfo {year} {2003})}\BibitemShut {NoStop}%
\bibitem [{\citenamefont {Kruis}\ \emph {et~al.}(1994)\citenamefont {Kruis}, \citenamefont {Schoonman},\ and\ \citenamefont {Scarlett}}]{kruis1994}%
  \BibitemOpen
  \bibfield  {author} {\bibinfo {author} {\bibfnamefont {F.~E.}\ \bibnamefont {Kruis}}, \bibinfo {author} {\bibfnamefont {J.}~\bibnamefont {Schoonman}}, \ and\ \bibinfo {author} {\bibfnamefont {B.}~\bibnamefont {Scarlett}},\ }\href@noop {} {\bibfield  {journal} {\bibinfo  {journal} {Journal of Aerosol Science}\ }\textbf {\bibinfo {volume} {25}},\ \bibinfo {pages} {1291} (\bibinfo {year} {1994})}\BibitemShut {NoStop}%
\bibitem [{spe()}]{specair}%
  \BibitemOpen
  \href@noop {} {}\bibinfo {note} {Laux, C.O., “Radiation and Nonequilibrium Collisional-Radiative Models,” von Karman Institute Lecture Series 2002-07, Physico-Chemical Modeling of High Enthalpy and Plasma Flows, eds. D. Fletcher, J.-M. Charbonnier, G.S.R. Sarma, and T. Magin, Rhode-Saint-Genèse, Belgium (2002).}\BibitemShut {Stop}%
\bibitem [{\citenamefont {Bruggeman}\ \emph {et~al.}(2014)\citenamefont {Bruggeman}, \citenamefont {Sadeghi}, \citenamefont {Schram},\ and\ \citenamefont {Linss}}]{bruggeman2014}%
  \BibitemOpen
  \bibfield  {author} {\bibinfo {author} {\bibfnamefont {P.~J.}\ \bibnamefont {Bruggeman}}, \bibinfo {author} {\bibfnamefont {N.}~\bibnamefont {Sadeghi}}, \bibinfo {author} {\bibfnamefont {D.}~\bibnamefont {Schram}}, \ and\ \bibinfo {author} {\bibfnamefont {V.}~\bibnamefont {Linss}},\ }\href@noop {} {\bibfield  {journal} {\bibinfo  {journal} {Plasma Sources Science and Technology}\ }\textbf {\bibinfo {volume} {23}},\ \bibinfo {pages} {023001} (\bibinfo {year} {2014})}\BibitemShut {NoStop}%
\bibitem [{\citenamefont {Nikhar}\ \emph {et~al.}(2020)\citenamefont {Nikhar}, \citenamefont {Rechenberg}, \citenamefont {Becker},\ and\ \citenamefont {Baryshev}}]{nikhar2020}%
  \BibitemOpen
  \bibfield  {author} {\bibinfo {author} {\bibfnamefont {T.}~\bibnamefont {Nikhar}}, \bibinfo {author} {\bibfnamefont {R.}~\bibnamefont {Rechenberg}}, \bibinfo {author} {\bibfnamefont {M.~F.}\ \bibnamefont {Becker}}, \ and\ \bibinfo {author} {\bibfnamefont {S.~V.}\ \bibnamefont {Baryshev}},\ }\href@noop {} {\bibfield  {journal} {\bibinfo  {journal} {Journal of Applied Physics}\ }\textbf {\bibinfo {volume} {128}} (\bibinfo {year} {2020})}\BibitemShut {NoStop}%
\bibitem [{\citenamefont {Agarwal}\ \emph {et~al.}(2012)\citenamefont {Agarwal}, \citenamefont {Rauf},\ and\ \citenamefont {Collins}}]{agarwal2012}%
  \BibitemOpen
  \bibfield  {author} {\bibinfo {author} {\bibfnamefont {A.}~\bibnamefont {Agarwal}}, \bibinfo {author} {\bibfnamefont {S.}~\bibnamefont {Rauf}}, \ and\ \bibinfo {author} {\bibfnamefont {K.}~\bibnamefont {Collins}},\ }\href@noop {} {\bibfield  {journal} {\bibinfo  {journal} {Plasma Sources Science and Technology}\ }\textbf {\bibinfo {volume} {21}},\ \bibinfo {pages} {055012} (\bibinfo {year} {2012})}\BibitemShut {NoStop}%
\bibitem [{\citenamefont {Godyak}\ \emph {et~al.}(1991)\citenamefont {Godyak}, \citenamefont {Piejak},\ and\ \citenamefont {Alexandrovich}}]{godyak1991}%
  \BibitemOpen
  \bibfield  {author} {\bibinfo {author} {\bibfnamefont {V.}~\bibnamefont {Godyak}}, \bibinfo {author} {\bibfnamefont {R.}~\bibnamefont {Piejak}}, \ and\ \bibinfo {author} {\bibfnamefont {B.}~\bibnamefont {Alexandrovich}},\ }\href@noop {} {\bibfield  {journal} {\bibinfo  {journal} {Journal of applied physics}\ }\textbf {\bibinfo {volume} {69}},\ \bibinfo {pages} {3455} (\bibinfo {year} {1991})}\BibitemShut {NoStop}%
\bibitem [{\citenamefont {Godyak}\ and\ \citenamefont {Piejak}(1990)}]{godyak1990}%
  \BibitemOpen
  \bibfield  {author} {\bibinfo {author} {\bibfnamefont {V.}~\bibnamefont {Godyak}}\ and\ \bibinfo {author} {\bibfnamefont {R.}~\bibnamefont {Piejak}},\ }\href@noop {} {\bibfield  {journal} {\bibinfo  {journal} {Physical review letters}\ }\textbf {\bibinfo {volume} {65}},\ \bibinfo {pages} {996} (\bibinfo {year} {1990})}\BibitemShut {NoStop}%
\bibitem [{\citenamefont {Moravej}\ \emph {et~al.}(2004)\citenamefont {Moravej}, \citenamefont {Yang}, \citenamefont {Nowling}, \citenamefont {Chang}, \citenamefont {Hicks},\ and\ \citenamefont {Babayan}}]{moravej2004}%
  \BibitemOpen
  \bibfield  {author} {\bibinfo {author} {\bibfnamefont {M.}~\bibnamefont {Moravej}}, \bibinfo {author} {\bibfnamefont {X.}~\bibnamefont {Yang}}, \bibinfo {author} {\bibfnamefont {G.}~\bibnamefont {Nowling}}, \bibinfo {author} {\bibfnamefont {J.}~\bibnamefont {Chang}}, \bibinfo {author} {\bibfnamefont {R.}~\bibnamefont {Hicks}}, \ and\ \bibinfo {author} {\bibfnamefont {S.}~\bibnamefont {Babayan}},\ }\href@noop {} {\bibfield  {journal} {\bibinfo  {journal} {Journal of applied physics}\ }\textbf {\bibinfo {volume} {96}},\ \bibinfo {pages} {7011} (\bibinfo {year} {2004})}\BibitemShut {NoStop}%
\bibitem [{\citenamefont {Raizer}(1997)}]{raizer1997}%
  \BibitemOpen
  \bibfield  {author} {\bibinfo {author} {\bibfnamefont {Y.~P.}\ \bibnamefont {Raizer}},\ }\href@noop {} {\emph {\bibinfo {title} {Gas discharge physics}}},\ edited by\ \bibinfo {editor} {\bibfnamefont {J.~E.}\ \bibnamefont {Allen}},\ Vol.~\bibinfo {volume} {2}\ (\bibinfo  {publisher} {Springer},\ \bibinfo {year} {1997})\BibitemShut {NoStop}%
\bibitem [{\citenamefont {Assael}\ \emph {et~al.}(2023)\citenamefont {Assael}, \citenamefont {Antoniadis}, \citenamefont {Velliadou},\ and\ \citenamefont {Wakeham}}]{assael2023}%
  \BibitemOpen
  \bibfield  {author} {\bibinfo {author} {\bibfnamefont {M.~J.}\ \bibnamefont {Assael}}, \bibinfo {author} {\bibfnamefont {K.~D.}\ \bibnamefont {Antoniadis}}, \bibinfo {author} {\bibfnamefont {D.}~\bibnamefont {Velliadou}}, \ and\ \bibinfo {author} {\bibfnamefont {W.~A.}\ \bibnamefont {Wakeham}},\ }\href@noop {} {\bibfield  {journal} {\bibinfo  {journal} {International Journal of Thermophysics}\ }\textbf {\bibinfo {volume} {44}},\ \bibinfo {pages} {85} (\bibinfo {year} {2023})}\BibitemShut {NoStop}%
\bibitem [{\citenamefont {Sun}\ and\ \citenamefont {Venart}(2005)}]{sun2005}%
  \BibitemOpen
  \bibfield  {author} {\bibinfo {author} {\bibfnamefont {L.}~\bibnamefont {Sun}}\ and\ \bibinfo {author} {\bibfnamefont {J.}~\bibnamefont {Venart}},\ }\href@noop {} {\bibfield  {journal} {\bibinfo  {journal} {International journal of thermophysics}\ }\textbf {\bibinfo {volume} {26}},\ \bibinfo {pages} {325} (\bibinfo {year} {2005})}\BibitemShut {NoStop}%
\bibitem [{\citenamefont {Forst}(1991)}]{forst1991}%
  \BibitemOpen
  \bibfield  {author} {\bibinfo {author} {\bibfnamefont {W.}~\bibnamefont {Forst}},\ }\href@noop {} {\bibfield  {journal} {\bibinfo  {journal} {The Journal of Physical Chemistry}\ }\textbf {\bibinfo {volume} {95}},\ \bibinfo {pages} {3612} (\bibinfo {year} {1991})}\BibitemShut {NoStop}%
\bibitem [{\citenamefont {Fedoseev}\ \emph {et~al.}(1979)\citenamefont {Fedoseev}, \citenamefont {Vnukov},\ and\ \citenamefont {Derjaguin}}]{fedoseev1979}%
  \BibitemOpen
  \bibfield  {author} {\bibinfo {author} {\bibfnamefont {D.}~\bibnamefont {Fedoseev}}, \bibinfo {author} {\bibfnamefont {S.}~\bibnamefont {Vnukov}}, \ and\ \bibinfo {author} {\bibfnamefont {B.}~\bibnamefont {Derjaguin}},\ }\href@noop {} {\bibfield  {journal} {\bibinfo  {journal} {Carbon}\ }\textbf {\bibinfo {volume} {17}},\ \bibinfo {pages} {453} (\bibinfo {year} {1979})}\BibitemShut {NoStop}%
\bibitem [{\citenamefont {Matsui}\ \emph {et~al.}(1987)\citenamefont {Matsui}, \citenamefont {Yuuki}, \citenamefont {Morita},\ and\ \citenamefont {Tachibana}}]{matsui1987}%
  \BibitemOpen
  \bibfield  {author} {\bibinfo {author} {\bibfnamefont {Y.}~\bibnamefont {Matsui}}, \bibinfo {author} {\bibfnamefont {A.}~\bibnamefont {Yuuki}}, \bibinfo {author} {\bibfnamefont {N.}~\bibnamefont {Morita}}, \ and\ \bibinfo {author} {\bibfnamefont {K.}~\bibnamefont {Tachibana}},\ }\href@noop {} {\bibfield  {journal} {\bibinfo  {journal} {Japanese journal of applied physics}\ }\textbf {\bibinfo {volume} {26}},\ \bibinfo {pages} {1575} (\bibinfo {year} {1987})}\BibitemShut {NoStop}%
\bibitem [{\citenamefont {Kang}\ and\ \citenamefont {Musgrave}(2000)}]{kang2000}%
  \BibitemOpen
  \bibfield  {author} {\bibinfo {author} {\bibfnamefont {J.~K.}\ \bibnamefont {Kang}}\ and\ \bibinfo {author} {\bibfnamefont {C.~B.}\ \bibnamefont {Musgrave}},\ }\href@noop {} {\bibfield  {journal} {\bibinfo  {journal} {The Journal of Chemical Physics}\ }\textbf {\bibinfo {volume} {113}},\ \bibinfo {pages} {7582} (\bibinfo {year} {2000})}\BibitemShut {NoStop}%
\bibitem [{\citenamefont {Haubner}(2021)}]{haubner2021}%
  \BibitemOpen
  \bibfield  {author} {\bibinfo {author} {\bibfnamefont {R.}~\bibnamefont {Haubner}},\ }\href@noop {} {\bibfield  {journal} {\bibinfo  {journal} {ChemTexts}\ }\textbf {\bibinfo {volume} {7}},\ \bibinfo {pages} {10} (\bibinfo {year} {2021})}\BibitemShut {NoStop}%
\bibitem [{\citenamefont {Bachmann}\ \emph {et~al.}(1991)\citenamefont {Bachmann}, \citenamefont {Leers},\ and\ \citenamefont {Lydtin}}]{bachmann1991}%
  \BibitemOpen
  \bibfield  {author} {\bibinfo {author} {\bibfnamefont {P.~K.}\ \bibnamefont {Bachmann}}, \bibinfo {author} {\bibfnamefont {D.}~\bibnamefont {Leers}}, \ and\ \bibinfo {author} {\bibfnamefont {H.}~\bibnamefont {Lydtin}},\ }\href@noop {} {\bibfield  {journal} {\bibinfo  {journal} {Diamond and related materials}\ }\textbf {\bibinfo {volume} {1}},\ \bibinfo {pages} {1} (\bibinfo {year} {1991})}\BibitemShut {NoStop}%
\end{thebibliography}%
\end{document}